\documentclass[aps,prb,amsmath,amssymb,footinbib,showpacs,twocolumn,superscriptaddress]{revtex4-1}
\usepackage{amsmath}
\usepackage{amssymb}
\usepackage{amsthm}
\usepackage{setspace}
\usepackage{graphicx}
\usepackage{braket}
\usepackage{mathrsfs}
\usepackage{float}
\usepackage[colorlinks = true,linkcolor = red,urlcolor  = blue,citecolor = blue,anchorcolor = blue]{hyperref}
\usepackage[utf8]{inputenc}
\usepackage[english]{babel}
\usepackage{bm}

\begin{document}

\title{Robust low-energy Andreev bound states in semiconductor-superconductor structures: Importance of partial separation of component Majorana bound states}

\author{Tudor D. Stanescu}
\affiliation{Department of Physics and Astronomy, West Virginia University, Morgantown, WV 26506, USA}

\author{Sumanta Tewari}
\affiliation{Department of Physics and Astronomy, Clemson University, Clemson, SC 29634, USA}

\begin{abstract}
Robust topologically trivial  low-energy Andreev bound states (ABSs) induced by position-dependent effective potentials have recently come under renewed focus in light of a remarkable set of experiments observing robust quantized zero-bias conductance plateaus in semiconductor-superconductor heterostructures.  We show that (1) the partial spatial separation of the wave functions of the component Majorana bound states (MBSs) is crucial for the creation and stability of topologically trivial near-zero-energy Andreev bound states, (2) the signs of the spin polarizations of the component MBSs can be either the same or opposite, depending on the profile of the inducing potential, and (3) the spin polarizations do not play a fundamental role in generating vastly different coupling strengths  to local probes and/or ensuring the robustness of the near-zero-energy ABS. Consequently, in contrast to recent theoretical claims (Vuik et al., arXiv:1806.02801), we find that a robust, quantized zero-bias conductance plateau of height $\sim 2e^2/h$ measured in the topologically trivial regime necessarily requires partially separated ABSs (ps-ABSs), independent of the relative signs of the spin-polarizations. In addition, we show that (4) well-defined energy splitting oscillations involve MBSs characterized by exponential tails pointing toward each other and that (5) ps-ABSs generated by the tunnel barrier itself produce zero-bias conductance peaks with a characteristic width that increases strongly with the applied magnetic field. Finally, we propose (6) a quantitative scheme for analyzing the stability of Majorana modes based on probability distributions of splitting susceptibilities and show that a ps-ABS mode can be remarkably robust when  judged based on its signature in a charge tunneling experiment, but, in essence, is topologically unprotected.
\end{abstract}
	
\maketitle

\section{Introduction}\label{SecI}
Semiconductor nanowires with proximity-induced superconductivity, strong Rashba spin-orbit coupling, and magnetic field applied parallel to the wire were predicted theoretically \cite{sau2010generic,tewari2010theorem,alicea2010majorana,sau2010non,lutchyn2010majorana,oreg2010helical,stanescu2011majorana} to support a pair of topologically-protected Majorana zero modes (MZMs) \cite{read2000paired,kitaev2001unpaired,nayak2008non,beenakker2013search,elliott2015colloquium} localized at the opposite ends of the wire. Owing to the interest stemming from its potential use as a platform for topological quantum computation (TQC),\cite{kitaev2001unpaired,nayak2008non} this system has been the beneficiary of tremendous experimental progress in the past few years. \cite{mourik2012signatures,deng2012anomalous,das2012zero,rokhinson2012fractional,churchill2013superconductor,finck2013anomalous,albrecht2016exponential,
deng2016majorana,zhang2017ballistic,chen2017experimental,nichele2017scaling,zhang2018quantized} One of the most recent important developments has been the observation of quantized zero-bias conductance plateaus in local charge tunneling experiments,\cite{zhang2018quantized} with the height of the plateau, which develops as a function of changing external parameters, such as Zeeman field and tunnel barrier height, equal to the  theoretically predicted height ($2e^2/h$) required by topological MZMs. \cite{sengupta2001midgap,akhmerov2009electrically,law2009majorana,flensberg2010tunneling} Numerous previous theoretical works on proximitized semiconductor nanowires with Rashba spin-orbit coupling and applied Zeeman field have shown the formation of zero-bias conductance peaks (ZBCPs) even in the absence of topological MZMs, due to disorder, \cite{bagrets2012class,liu2012zero,degottardi2013majoranaprl,degottardi2013majoranaprb,rainis2013towards,adagideli2014effects} non uniform system parameters, \cite{kells2012near,chevallier2012mutation,roy2013topologically,san2013multiple,ojanen2013topological,stanescu2014nonlocality,cayao2015sns,klinovaja2015fermionic,san2016majorana,fleckenstein2017decaying} weak antilocalization, \cite{pikulin2012zero} and coupling to a quantum dot.\cite{prada2012transport,liu2017andreev} 
However, most of these peaks of non-topological origin do not typically result in a $2e^2/h$-quantized conductance plateau, with the height of the plateau remaining unchanged against variations of the control parameters, such as Zeeman field and tunnel barrier height, in spite of some of the peaks exhibiting fairly robust pinning near zero energy.  Consequently, in the recent experiments \cite{zhang2018quantized} the observation of zero-bias conductance peaks of height $2e^2/h$ that generate quantized conductance plateaus with the variation of the control parameters, a feature typically associated with the existence of isolated topological MZMs, has been used as key evidence for the presence of non-Abelian, topologically protected MZMs localized at the opposite ends of the experimental system.

On the other hand, in a recent theoretical work \cite{Moore-recent} it has been shown that quantized conductance plateaus of height $2e^2/h$ (which are robust over large ranges of Zeeman field and tunnel barrier potential) are possible in a topologically trivial system due to the presence of low-energy Andreev bound states (ABSs) whose component Majorana bound states (MBSs) are somewhat shifted in space: the so-called partially separated ABSs (ps-ABS) introduced in Ref.~[\onlinecite{moore2018two-terminal}]. 
The ps-ABSs are topologically trivial, being characterized by energy splittings due to the overlap of the component MBSs that are sensitively dependent on local quantities, such as the local values of the spin-orbit coupling and the slope of the quantum dot potential. Moreover, the separation of the component MBSs cannot be controlled externally \cite{moore2018two-terminal} if the ps-ABSs are generated ``accidentally''  by some inhomogeneity and would be hard to control even when the ps-ABS is intentionally produced (e.g., one would have to control the slope of an effective potential that has a complicated, hard-to-determine relation to an applied gate voltage). Consequently, ps-ABSs do not provide a natural  platform for building a topologically protected qubit.\cite{Karzig2017}

Despite the topologically trivial nature of ps-ABSs, the emergence of quantized zero-bias conductance plateaus from local tunneling off ps-ABSs can be naturally understood based on the real-space properties of the component MBSs.
Essentially, due to the partial spatial separation of the component MBSs, when one couples locally to the end of a wire, one couples  strongly to only one of the constituent MBSs of a ps-ABS, while the other component remains practically ``invisible.'' Since ps-ABSs can be produced rather generically by local potentials induced by, e.g., tunnel gates in a quantum dot associated with the uncovered segment of a nanowire, the observation of quantized conductance plateaus in local charge tunneling experiments \textit{cannot} be taken as clinching evidence for topological MZMs. 
In a recent work\cite{vuik2018reproducing} it is argued that a semiconductor-superconductor (SM-SC) heterostructures with smooth confinement potential at the end, similar to the system studied in Ref.~[\onlinecite{kells2012near}], can support robust near-zero-energy ABSs consisting of {\em fully overlapping} component MBSs. In this scenario, the so-called quasi-Majoranas (i.e., the MBS components of the low-energy ABS) experience different effective barrier potentials to the external lead in the presence of a sufficiently strong Zeeman field as a result of being associated with different spin-split sub-bands (i.e., having different spin polarizations). Thus, the overlapping quasi-Majoranas produce a quantized conductance plateau of height $2e^2/h$ if the coupling to the external lead of one component  is strongly suppressed with respect to that of the component with opposite spin polarization. In this work we examine this mechanism that has been proposed for explaining the emergence of quantized plateaus in topologically trivial systems by studying systematically the real space and spin properties of the component Majorana wave functions associated with low-energy ABSs  emerging in nonhomogeneous systems, including ABSs capable of generating robust zero-bias conductance plateaus in a local charge tunneling experiment.  

To examine these questions, we first identify two basic mechanisms for the emergence of low-energy ABSs that are associated with variations of the effective potential due to, e.g., tunnel gates in the quantum dot region and/or a position-dependent work function difference between the semiconductor and the superconductor. Based on extensive numerical calculations, we show that  in both cases the partial spatial separation of the wave functions of the component Majorana bound states is crucial for the emergence of robust topologically trivial near-zero-energy Andreev bound states at values of the magnetic field less than the critical value corresponding to the topological phase transition. This partial spatial separation of the component Majorana wave functions can be induced by monotonic  potentials with  finite average slope (which can lead to the partial separation of the intrinsic sub-gap ABSs and their collapse to near-zero energy), as well as by non-monotonic valley or hill potentials (see Sec. \ref{SecII}).  We emphasize that  the partial-separation mechanisms discussed in this work can be active in (effectively) single-band systems (i.e., systems with well-separated, uncoupled confinement-induced bands, where the low-energy physics is controlled by the topmost occupied band), as well as multi-band systems (i.e., systems characterized by multi-band occupancy and strong inter-band coupling). In addition, multi-band systems can support an alternative  mechanism that pins trivial ABS states near zero energy, the so-called  inter-band coupling mechanism.\cite{Woods2019}  In this case, band repulsion resulting from the coupling of two or more confinement-induced low-energy bands can pin the ABS near zero energy over a significant range of control parameters (e.g., Zeeman field), even though its Majorana components are not separated spatially.\cite{Woods2019} By contrast, the mechanisms investigated in this work rely critically on the partial spatial separation of the component MBSs. 

We find that the signs of the spin polarizations of the component MBSs comprising a low-energy ABS can be equal or opposite to each other, depending on the profile of the potential that induces them. For example, in the case of  smooth confinement potentials the spin polarizations of the two MBSs  are different (see, e.g., Fig. \ref{FG2}), while for non-monotonic valley-like or hill-like potentials they are the same (see Figs. \ref{FG6}--\ref{FG8}). We also find that, in general, the spin polarizations of the component MBSs do not play a fundamental role in generating vastly different coupling strengths to local probes and/or ensuring the robustness of the near-zero-energy ABS (see Sec. \ref{SecIIIB}, e.g., Fig. \ref{FG14}, Figs. \ref{FG16}--\ref{FG18}, or Figs. \ref{FG20}--\ref{FG22}). Furthermore, the component MBSs of a low-energy ABS may couple differently to a local probe because of the difference in the characteristic wave numbers. However, in the absence of a partial spatial separation of the wave functions, the energy of the corresponding ABS remains comparable to the bulk gap (see, e.g., Figs. \ref{FG10} and \ref{FG12}) and does not  produce a zero-bias conductance peak in local charge tunneling. Thus, we conclude that the emergence of robust, quantized, zero-bias conductance plateaus of height $\sim 2e^2/h$ in local charge tunneling experiments in the topologically trivial regime of SM-SC nanowire heterostructures necessarily requires ps-ABSs whose component MBSs are spatially separated by smooth confinement potential and/or potential hills, independent of the relative signs of the spin-polarizations. We also show that low-energy states characterized by well-defined energy splitting oscillations involve MBSs having exponential tails that point toward each other [see Figs. \ref{FG8} and \ref{FG5}(c) and Figs. \ref{FG20}--\ref{FG22}]. To experimentally demonstrate energy-splitting oscillations in short wires [for values of the magnetic field above the critical field corresponding to the topological quantum phase transition (TQPT)] one has to ensure hard confinement and the absence of (unwanted) quantum dots at the ends of the wire, which may suppress the splitting oscillations.
Finally, we show that the ps-ABSs induced by soft confinement have a strong characteristic signature in charge tunneling experiments: the width of the ZBCP increases monotonically with the applied Zeeman field as a result of the effective tunnel barrier being field-dependent (see Fig. \ref{FG25}). 

The remainder of this paper is organized as follows. In Sec. \ref{SecII} we introduce the basic types of ps-ABSs and we summarize their main real-space and spin properties.  Section \ref{SecIII} focuses on the role of the spatial separation of the component MBSs in driving the collapse of ps-ABSs toward zero energy. In particular, we show that the ps-ABSs generated by soft confinement are adiabatically connected to the intrinsic ABSs that emerge generically in clean wires with finite chemical potential. We also investigate the spin structure of ps-ABSs  showing that the spin does not play a fundamental role in their collapse to zero energy.  Some specific charge tunneling signatures of ps-ABSs are discussed in Sec.  \ref{SecIV}. The  robustness of ps-ABSs in the presence of disorder is investigated in Sec. \ref{SecV}. We conclude in Sec. \ref{SecVI} with a summary of the main results and with our conclusions. 

\section{Basic types of potential-induced low-energy Andreev bound states}\label{SecII}

It is commonly believed that Andreev bound states (ABSs) emerging  in hybrid semiconductor-superconductor devices can have near-zero energy
in the topologically trivial regime as long as the confinement potential is sufficiently smooth.\cite{kells2012near}
By contrast, in this section we show that there are two different basic mechanisms responsible for the emergence of topologically trivial low-energy ABSs in effectively single-band systems with inhomogeneous potential. These mechanisms are associated with  ($\alpha$) shore potential regions with finite average slope and ($\beta$) ``nearly dry'' potential wells/``almost submerged'' potential hills, where the ``water level''  is given by the chemical potential.   The smooth confinement potential is a particular case of the ``shore potential'' scenario.

Before we discuss the specific theoretical models for the SM-SC nanowire heterostructures with position-dependent effective potential,  let us first identify the relevant characteristic length scales. Consider a hybrid system with position-dependent effective potential $V(x)$, chemical potential $\mu$, and (half) Zeeman splitting $\Gamma$; we define the high/low ``water levels'' $x_\pm^{(i)}$ as the solutions of the equation
\begin{equation}
V(x_\pm) \pm \Gamma = \mu.  \label{Vpm}
\end{equation}
The characteristic ``shore width'' associated with scenario ($\alpha$) is $L^*(\Gamma) = |x_+(\Gamma) - x_-(\Gamma)|$. Of course, the magnitude of $L^*$ is determined by the Zeeman splitting and by the {\em average} slope of the potential over the ($\Gamma$-dependent) shore region. In the case of  a potential well (hill) there are two $x_-$ ($x_+$) solutions (with no solution for the opposite spin-split sub-band)  and the corresponding characteristic length scale is $L^*(\Gamma) = x_-^{(2)} - x_-^{(1)}$ [$L^*(\Gamma) = x_+^{(2)} - x_+^{(1)}$]. Low-energy ABS modes collapse toward zero energy when $L^*$ becomes comparable with a certain characteristic length scale of the ABS  Majorana components, as shown below. Note that obtaining a large enough value of $L^*$ does not involve any ``smoothness'' requirement.

In this work we focus on the single-band approximation, which is valid when the occupation is low and the inter-band spacing is large compared to other energy scales in the problem. In this approximation the occupied bands can be treated as being independent (i.e., not coupled to one another) and the low-energy physics of the hybrid SM-SC system can be captured using a single-band model. Specifically, we consider the simple effective tight-binding model given by the Bogoliubov--de Gennes (BdG) Hamiltonian
\begin{eqnarray}
H &=& -t\sum_{\langle i, j\rangle, \sigma} c_{i\sigma}^\dagger c_{j\sigma} +\sum_{i, \sigma}(V_i - \mu)  c_{i\sigma}^\dagger c_{i\sigma} +\Gamma\sum_i c_i^\dagger \sigma_x c_i \nonumber \\
&+&i\frac{\alpha}{2}\sum_{\langle i, j\rangle}\left( c_i^\dagger \sigma_y c_j + {\rm H.c.}\right) + \Delta\sum_i\left( c_{i \uparrow}^\dagger c_{i\downarrow} + {\rm H.c.}\right),   \label{H}
\end{eqnarray}
where $\langle i, j\rangle$ are nearest-neighbor sites in a one-dimensional lattice, $c_i^\dagger =(c_{i \uparrow}^\dagger, c_{i \downarrow}^\dagger)$ is the electron creation operator on site $i$, and $\sigma_\nu$ (with $\nu = x, y, z$) are Pauli matrices. The model parameters are the nearest-neighbor hopping $t$, the chemical potential $\mu$, the (half) Zeeman splitting $\Gamma$, the Rashba spin-orbit coupling $\alpha$, and the proximity-induced pairing $\Delta$. We assume the presence of a position-dependent effective potential $V_i = V(i a-i_0 a)$, where $a$ is the lattice constant and $i_0$ indexes the origin of the coordinate axis, $x=0$. The values of the model parameters used in the numerical calculations are $t=7.62~$meV, $\alpha=3~$meV, and $\Delta=0.25~$meV, unless specified otherwise. We study numerically the dependence of the low-energy states on the Zeeman field for different values of the chemical potential and different effective potential profiles. Note that, for a value of the lattice constant $a=10~$nm, the model parameters correspond to an effective mass $m=0.05m_0$ and a Rashba coefficient $\alpha_R =300~$meV\AA.  The effective mass is larger than the typical values of the effective mass in semiconductor nanowires (e.g., $m=0.023m_0$ in InAs) to mimic the effect of proximity-induced  energy renormalization.\cite{Stanescu2017a} Also note that we consider Zeeman splittings  $\Gamma$ up to $2-3~$meV, which correspond to magnetic fields on the order of $1.5-2.5~$T for a wire with a Land\'{e} g-factor $g\approx 20$. We emphasize that the topologically trivial features that represent the focus of our work typically emerge at significantly lower fields. 

\begin{figure}[t]
\begin{center}
\includegraphics[width=0.49\textwidth]{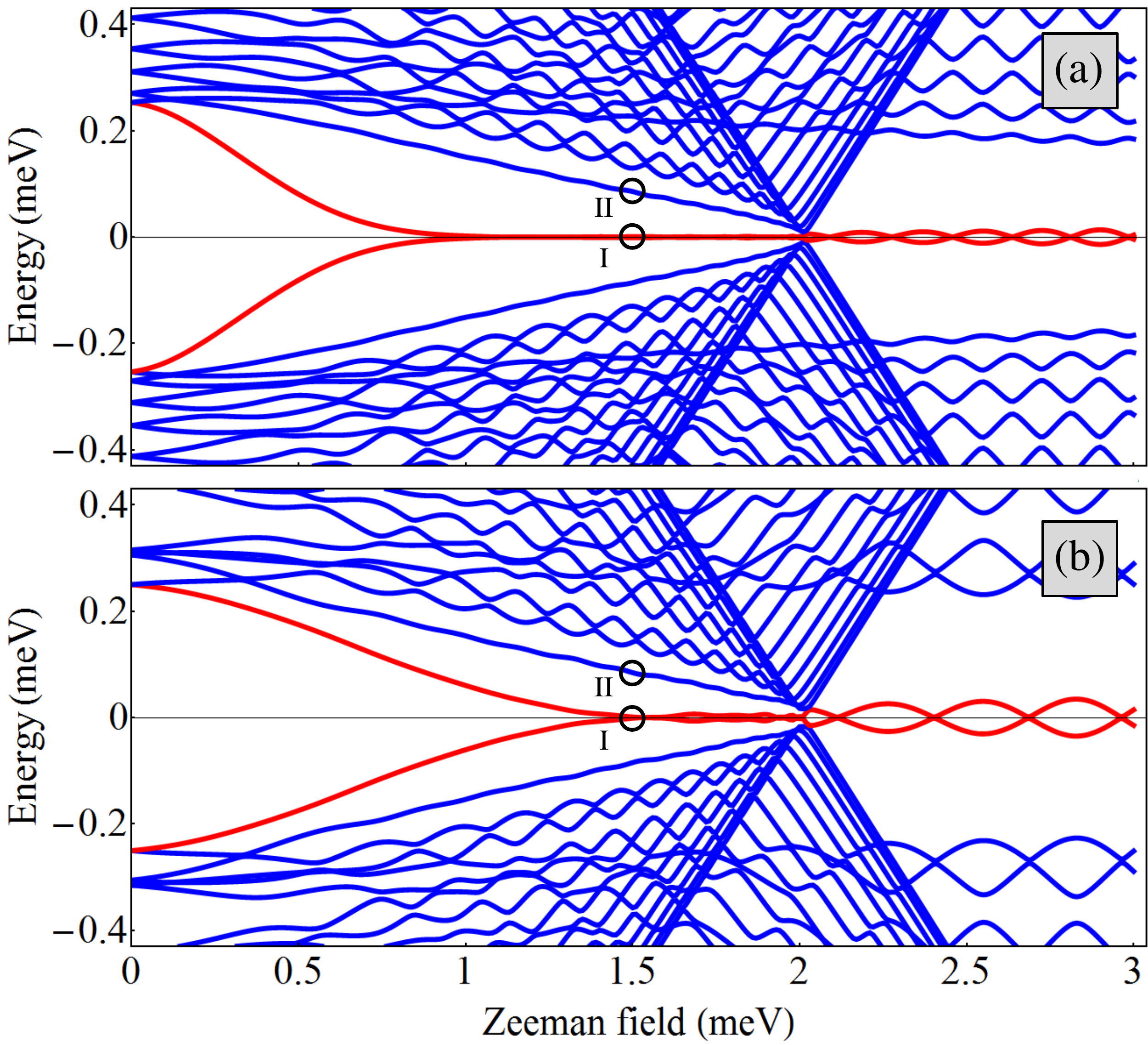}
\end{center}
\caption{Dependence of the low-energy spectrum on the applied Zeeman field for a system with effective potential given by Eq. (\ref{Vx1}) and chemical potential $\mu=2~$meV. The slope of the potential at the left end of the wire is (a) $\kappa = 9~$meV$/\mu$m and (b)  $\kappa = 36~$meV$/\mu$m. The bulk gap has a minimum at $\Gamma_c\approx\mu=2~$meV, the critical field associated with the TQPT. Note that the lowest-energy ABS mode localized near the left end of the wire collapses to zero energy at values of the Zeeman field $\Gamma<\Gamma_c$ (i.e., in the topologically trivial regime). The second-lowest-energy mode corresponds to an {\em intrinsic} ABS localized near the right end of the wire. The states marked by ``I'' and ``II'' are shown in Figs. \ref{FG2}--\ref{FG4}.}
\label{FG1}
\end{figure}

\subsection{Finite width potential shores}\label{SecIIA}

We first consider the emergence of low-energy ABSs within scenario ($\alpha$), i.e., potential shores with finite (average) slope. For concreteness,  we first assume that the system is characterized by a linear confining potential at the left end and a hard-wall confinement at the right end of the wire. Specifically, we have
\begin{equation}
V(x) = \left\{
\begin{array}{l}
-\kappa x  ~~~~~~~~~ {\rm if~}~x<0, \\
~~0 ~~~~~~~~~~~ {\rm if~}~0 \leq x \leq L, \\
~\infty ~~~~~~~~~~\!~{\rm if~}~x> L,
\end{array}\right.    \label{Vx1}
\end{equation}
where $L$ is the length of a (long) wire segment characterized by constant effective potential and $\kappa$ is the slope of the confining potential at the left end of the system.

\begin{figure}[t]
\begin{center}
\includegraphics[width=0.49\textwidth]{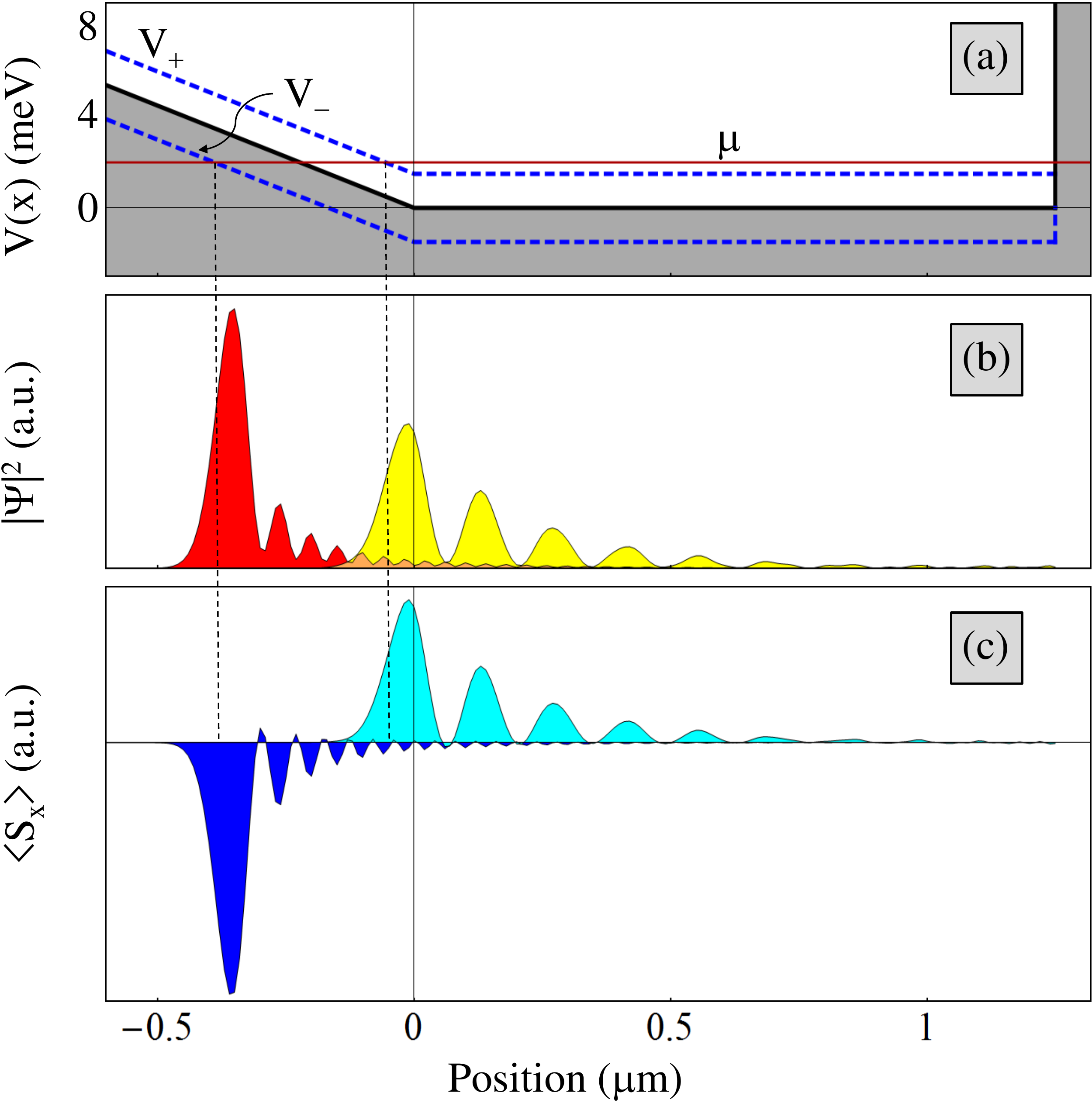}
\end{center}
\caption{(a) Potential profile corresponding to Eq. (\ref{Vx1}) with $\kappa=9~$meV$/\mu$m (shaded black line). The dashed blue lines correspond to $V_\pm(x) = V(x) \pm\Gamma$, where $\Gamma=1.5~$meV is the Zeeman field. The chemical potential $\mu=2~$meV (red line) intersects the $V_\mp$ profiles at points $x_-$ and $x_+$, respectively, given by Eq. (\ref{Vpm}). (b) Majorana wave functions [given by Eqs. (\ref{chiA} and \ref{chiB})] associated with the near-zero energy ABS marked ``I'' in Fig. \ref{FG1}(a). c) Spin density of the Majorana modes shown in panel (b).  Note that the main peaks of the MBS wave functions are localized near the $x_-$ and $x_+$ points, respectively. The corresponding  $x$-components of the spin density have opposite signs, revealing the fact that two Majoranas are associated with {\em different} spin-split sub-bands.}
\label{FG2}
\end{figure}

\begin{figure}[t]
\begin{center}
\includegraphics[width=0.49\textwidth]{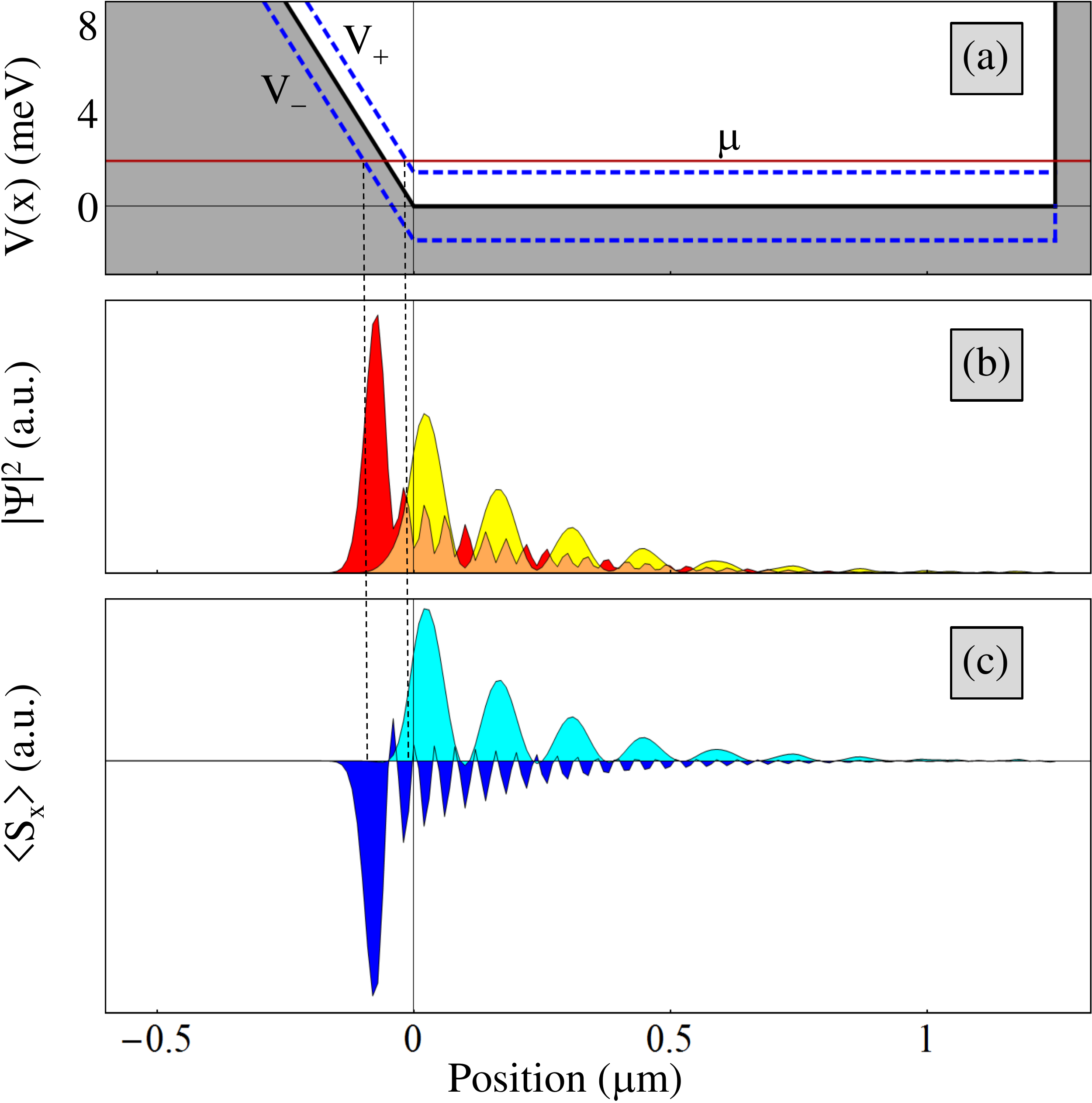}
\end{center}
\vspace{-3mm}
\caption{(a) Potential profile corresponding to Eq. (\ref{Vx1}) with $\kappa=36~$meV$/\mu$m (shaded black line) and the associated $V_\pm$ profiles (dashed blue lines).  (b) Majorana wave functions $\psi_A$ (red) and $\psi_B$ (yellow) [given by Eqs. (\ref{chiA} and \ref{chiB})] associated with the near-zero energy ABS marked I in Fig. \ref{FG1}(b). The orange area corresponds to the overlap of the two-component MBSs. c) Spin density of the Majorana modes shown in (b).}
\label{FG3}
\end{figure}

Figure \ref{FG1} shows two examples of low-energy ABS modes emerging in the topologically trivial regime in a system with smooth confinement potential given by Eq. (\ref{Vx1}) corresponding to two different values of the slope $\kappa$. By comparing panels (a) and (b) it is clear that the crossover Zeeman field $\Gamma_c^* < \Gamma_c$ associated with the collapse to zero energy of the ABS mode  increases with the slope $\kappa$ of the confining potential. In the trivial regime, the second-lowest-energy mode corresponds to an {\em intrinsic} sub-gap ABS localized near the right end of the wire, which is characterized by a hard-wall confining potential. It is also important to note that for $\Gamma > \Gamma_c$ the Majorana energy splitting oscillations in Fig. \ref{FG1}(b) are larger than those in panel (a) as a result of the {\em effective} length of the wire being shorter for large $\kappa$ (i.e., harder confinement, see also Figs. \ref{FG2} and \ref{FG3}).

A generic eigenstate of the BdG Hamiltonian (\ref{H}) can be expressed as a sum of two Majorana modes.
 Consider a low-energy solution $\phi_\epsilon$ corresponding to a positive energy $\epsilon \ll \Delta$ with a wave function (in the spinor representation) $\phi_\epsilon(i) = (u_{i\uparrow},   u_{i\downarrow}, v_{i\uparrow},   v_{i\downarrow})^T$. Particle-hole symmetry ensures the existence of a negative-energy solution of the BdG equation described by the wave function $\phi_{-\epsilon}(i) = (v_{i\uparrow}^*,   v_{i\downarrow}^*, u_{i\uparrow}^*,   u_{i\downarrow}^*)^T$. Using these solutions, we construct the linear combinations
\begin{eqnarray}
\psi_A(i) &=& \frac{1}{\sqrt{2}}\left[\phi_{\epsilon}(i) +\phi_{-\epsilon}(i)\right], \label{chiA}\\
\psi_B(i) &=& \frac{i}{\sqrt{2}}\left[\phi_{\epsilon}(i) -\phi_{-\epsilon}(i)\right].     \label{chiB}
\end{eqnarray}
These states have a spinor structure of the form $\psi_\alpha(i)=(\widetilde{u}_{\alpha i\uparrow}, \widetilde{u}_{\alpha i\downarrow}, \widetilde{u}_{\alpha i\uparrow}^*, \widetilde{u}_{\alpha i\downarrow}^*)^T$, where $\alpha = A, B$ and $u_{A,i,\sigma}=u_{i\sigma} +v_{i\sigma}^*$, while $u_{B,i,\sigma}=i(u_{i\sigma} -v_{i\sigma}^*)$, which manifestly satisfies the Majorana condition. We note that the Majorana representation of the eigenstates of the BdG Hamiltonian, $\phi_{\pm\epsilon} =  \frac{1}{\sqrt{2}}(\psi_A\pm i\psi_B)$
  is generic, but $\psi_A$ and $\psi_B$ are not eigenstates of $H$, except for $\epsilon = 0$, and we have $\langle \psi_\alpha |H|\psi_\alpha\rangle = 0$ and $\langle \psi_A |H|\psi_B\rangle = i \epsilon$. Note that using the Majorana basis allows a simple physical interpretation of low-energy ps-ABSs as (partially) overlapping Majorana modes emerging at the ends of a (typically short) segment of the wire where the topological condition is locally satisfied (i.e., within shore potential regions, nearly dry potential well bottoms, and almost submerged potential hill tops). In addition, using the Majorana basis and the concept of ps-ABS has practical relevance, as it interpolates continuously between a purely local ABS (consisting of two completely overlapping MBSs) and a pair of well-separated MBSs capable of supporting topological quantum computation. In this language, the critical problem is to determine the necessary separation and realize it in controllable devices.

The wave functions $\psi_A$ and $\psi_B$ corresponding to the  near-zero energy ABS marked `I' in Fig. \ref{FG1}(a) are represented in Fig. \ref{FG2}(b) as the red and yellow lines, respectively.
We also define the ($\nu$-component) of the spin-density as
\begin{equation}
\langle S_\nu\rangle(i) = \frac{1}{2}\sum_{s, s^\prime} \widetilde{u}_{is}^*[\sigma_\nu]_{s s^\prime}\widetilde{u}_{is^\prime}.
\end{equation}
Note that $\langle S_\nu\rangle$, which represents the particle component of the total spin density, can be probed using spin-resolved tunneling spectroscopy.
In Fig. \ref{FG2}(c), we show that the Majorana components of a low-energy ABS induced by a finite width potential shore belong to {\em different} spin-split sub-bands, as demonstrated by the opposite signs of the spin density in Fig. \ref{FG2}(c) [see also Fig. \ref{FG3}(c)].
It is also important to note that in this case the exponential tails of the Majorana wave functions point in the same direction (toward decreasing potential).

In general, we find that the component MBSs are characterized by two length scales: the width $\delta_M$ of the main peak and the characteristic length scale of the (exponentially decaying) envelope, $\xi_M$. The condition for the collapse of the energy of the sub-gap ABS mode to near-zero energy is
that the separation $L^* = |x_+-x_-|$ of the component MBSs, which is controlled by the width of the potential shore, should be larger than the width of the main MBS peak. Physically, one can view the potential shore as a ``locally topological'' segment that supports two (partially-overlapping) MBSs at its ends. The collapse toward zero energy is associated with this ``topological'' segment being longer than $\delta_M$. In terms of Zeeman splitting, the condition becomes $\Gamma>\Gamma_c^*$, where the crossover Zeeman field is given by the condition $L^*(\Gamma_c^*) = |x_+-x_-|\sim \delta_M$.

\begin{figure}[t]
\begin{center}
\includegraphics[width=0.49\textwidth]{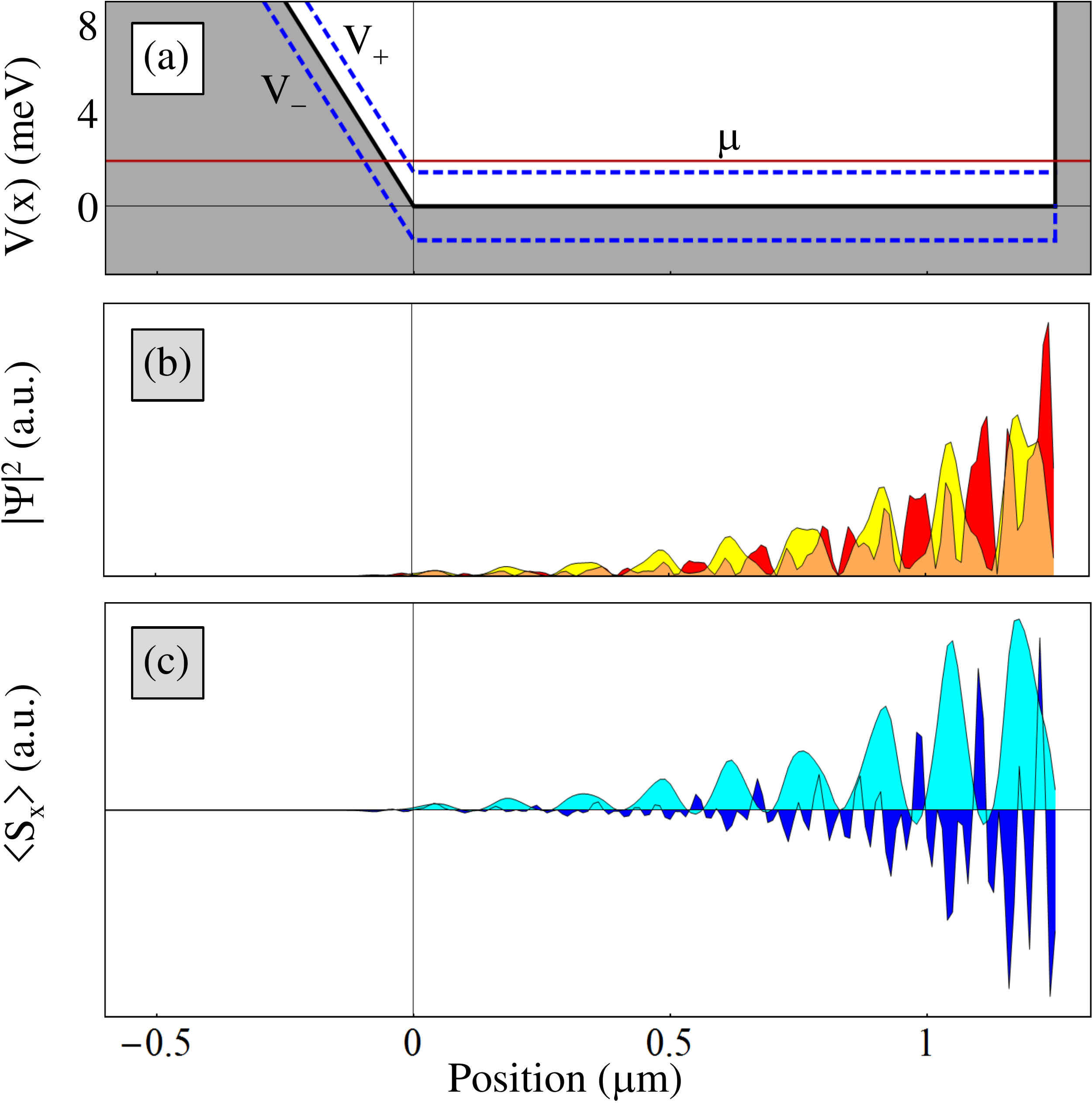}
\end{center}
\vspace{-3mm}
\caption{(a) Same as panel (a) in Fig. \ref{FG3}. (b) Majorana components of the intrinsic ABS marked `II' in Fig. \ref{FG1}. The ABS is localized at the right end of the system, which has hard-wall confinement, and does not depend on the details of the soft confinement potential at the left end. (c) Spin density of the Majorana modes shown in (b).}
\label{FG4}
\end{figure}

A direct consequence of these observations is that the crossover field $\Gamma_c^*$ increases with the (average) slope of the potential within the ``shore'' region. Also, for a given value of the Zeeman field, the separation of the component MBSs, which is given by $L^*$, decreases with the average slope. These properties are illustrated by the comparison between panels (a) and (b) in Fig. \ref{FG1}, which correspond to $\kappa=9~$meV$/\mu$m and $\kappa=36~$meV$/\mu$m, respectively, and by the comparison between the states marked `I' in these panels, which are shown in Figs. \ref{FG2} and \ref{FG3}, respectively. Note also that the energy of the potential-induced ABS ``sticks'' to zero (as long as $L^* > \delta_M$) despite the substantial overlap of the ``yellow MBS'' with the tail of the ``red MBS'' [orange areas in Figs. \ref{FG2}(b) and \ref{FG3}(b)].

It has been shown previously\cite{Huang2018a} that clean superconducting spin-orbit-coupled nanowires are generically characterized by finite energy in-gap Andreev bound states emerging below the topological quantum phase transition in systems with finite values of the chemical potential. We dub these low-energy states emerging in uniform systems (with hard confinement) as {\em intrinsic} Andreev bond states (i-ABS).
The Majorana components of the i-ABS  localized at the right end of a wire with effective potential described by Eq. (\ref{Vx1}) are shown in Fig. \ref{FG4}(b).  Unlike the component MBSs associated with the ps-ABS localized at the left end of the wire (which has smooth confinement), these Majorana modes are not separated. Consequently, the lowest-energy ABS modes illustrated in Figs. \ref{FG2} and\ref{FG3} can be interpreted as {\em extrinsic} ABS modes characterized by Majorana components partially separated (spatially)  by the position-dependent effective potential in the presence of a finite Zeeman field.

\subsection{Almost submerged potential hills and nearly dry potential wells}\label{SecIIB}

As an example of low-energy ABSs generated within scenario ($\beta$) ( i.e.,  by ``nearly dry'' potential wells or ``almost submerged'' potential hills, with the ``water level'' provided by the chemical potential) we first consider the low-energy modes of a system with position-dependent step-like potential defined by
\begin{equation}
V(x) = \left\{
\begin{array}{l}
V_0  ~~~~~~~~~\!~ {\rm if~}~|x|<d_{\rm v}, \\
~0 ~~~~~~~~~~ {\rm if~}~|x|>d_{\rm v}.
\end{array}\right.    \label{Vx2}
\end{equation}
When $V_0>0$ Eq. (\ref{Vx2}) describes a potential hill, while $V_0<0$ corresponds to a rectangular potential well.
Note that for values of the chemical potential near the top of the hill (or the bottom of the well), $\mu \sim V_0$, and a Zeeman field $\Gamma > |\mu-V_0|$, Eq.  (\ref{Vpm}) has solutions $x_+^{(2)}= -  x_+^{(1)}=d_{\rm v}$ ($x_-^{(2)}= -  x_-^{(1)}=d_{\rm v}$).

\begin{figure}[t]
\begin{center}
\includegraphics[width=0.49\textwidth]{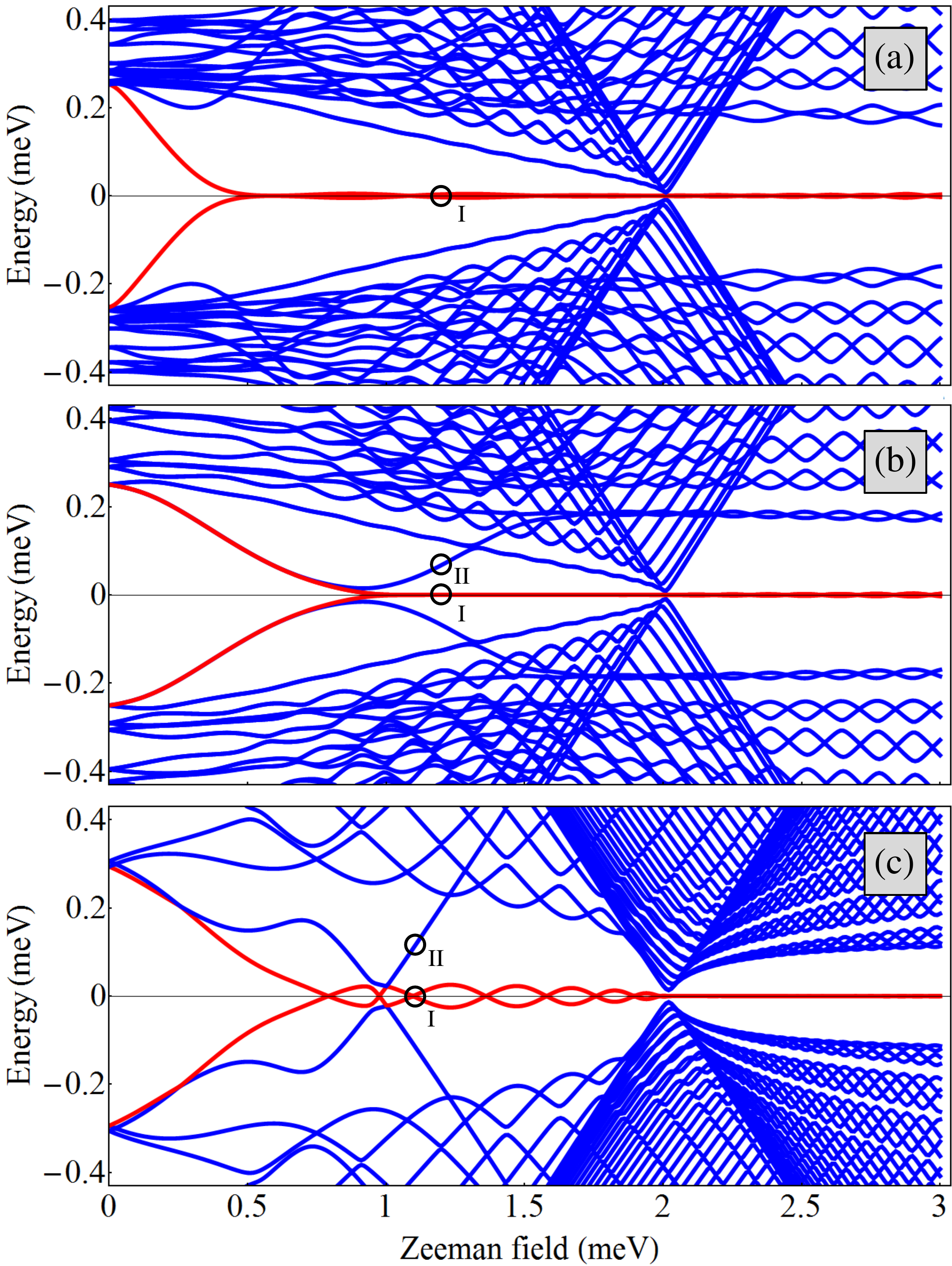}
\end{center}
\vspace{-3mm}
\caption{(a) Dependence of the low-energy spectrum on the applied Zeeman field for a system with chemical potential $\mu=2~$meV and effective step-like potential given by Eq. (\ref{Vx2}) with $V_0=2.1~$meV and $d_{\rm v}=0.25~\mu$m. (b) Same as in panel (a) for a Gaussian potential hill given by Eq. (\ref{Vx3}) with $V_0^\prime=3.2~$meV and $\delta_{\rm v}=0.23~\mu$m. (c) Low-energy spectrum for a Gaussian potential well given by Eq. (\ref{Vx3}) with $V_0^\prime=-3.2~$meV and $\delta_{\rm v}=0.23~\mu$m; the chemical potential is $\mu=-2~$meV.
The bulk gap has a minimum at $\Gamma_c=|\mu|=2~$meV, the critical field associated with the TQPT. Note that the lowest-energy ABS mode localized near the potential hill/well (red line) collapses to zero energy at values of the Zeeman field $\Gamma<\Gamma_c$ (i.e., in the topologically trivial regime). Also note that the potential well [panel (c)] induces large energy splitting oscillations, in contrast to the potential hill [(a) and (b)]. The states marked by ``I'' and ``II'' are shown in Figs. \ref{FG6}--\ref{FG8}.}
\label{FG5}
\end{figure}

\begin{figure}[t]
\begin{center}
\includegraphics[width=0.49\textwidth]{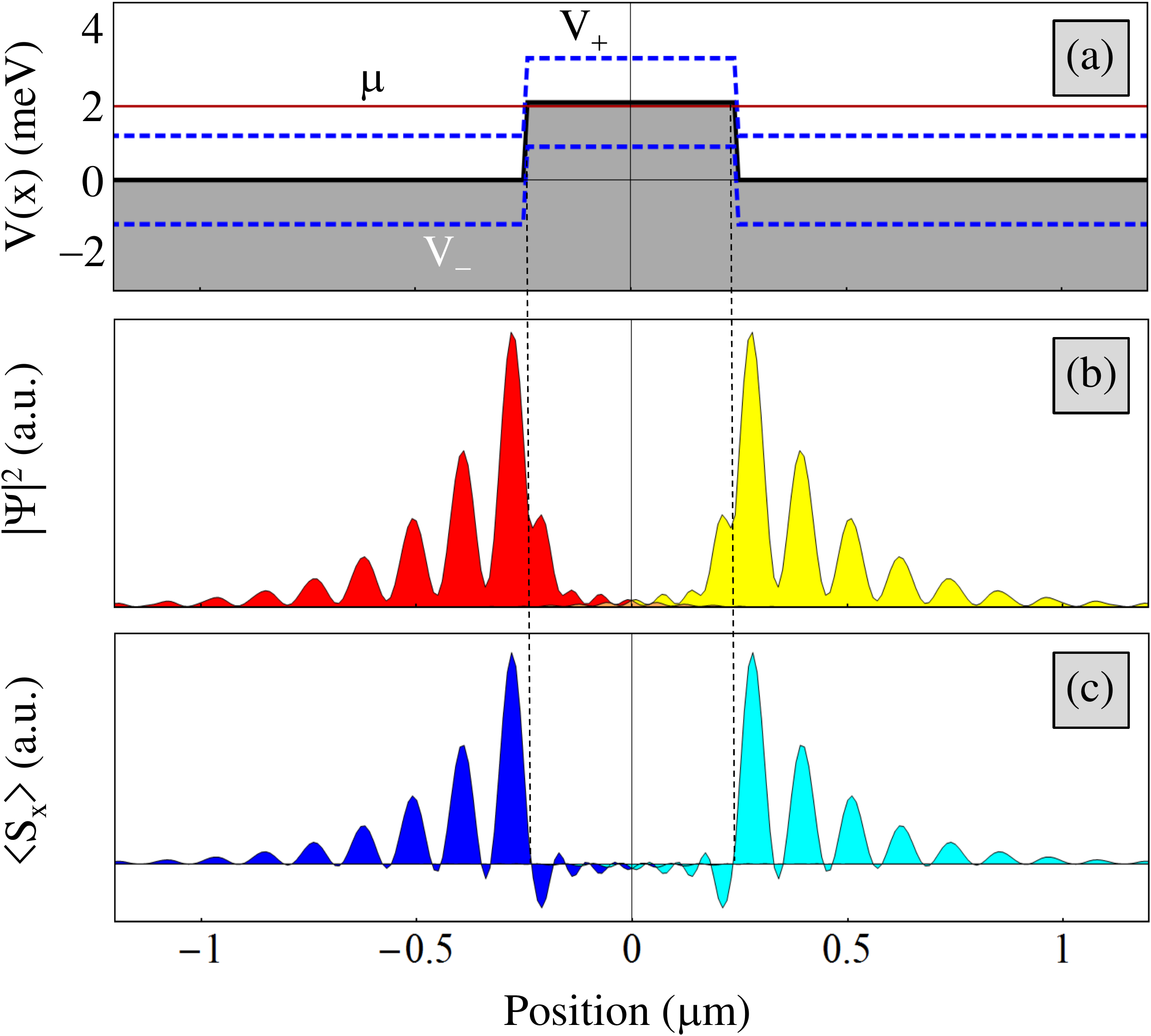}
\end{center}
\caption{(a) Potential profile corresponding to Eq. (\ref{Vx2}) with  $V_0=2.1~$meV and $d_{\rm v}=0.25~\mu$m (shaded black line). The dashed blue lines correspond to $V_\pm(x) = V(x) \pm\Gamma$, where $\Gamma=1.5~$meV is the Zeeman field. The chemical potential $\mu=2~$meV (red line) intersects the $V_+$ profile at $x_+^{(1)}=-d_{\rm v}$ and $x_+^{(2)}=d_{\rm v}$. (b) Majorana wave functions [given by Eqs. (\ref{chiA} and \ref{chiB})] associated with the near-zero energy ABS marked `I' in Fig. \ref{FG5}(a). (c) Spin density of the Majorana modes shown in panel (b).  Note that the main peaks of the MBS wave functions are localized near  $x_+^{(1)}$ and  $x_+^{(2)}$. The corresponding  $x$-components of the spin density have the same sign, revealing the fact that two Majoranas are associated with the {\em same} spin-split sub-band.}
\label{FG6}
\end{figure}

We also consider a ``smooth'' Gaussian  hill/well defined by
\begin{equation}
V(x) = V_0^\prime \exp\left(-\frac{x^2}{\delta_{\rm v}^2}\right). \label{Vx3}
\end{equation}
The purpose of studying this case is twofold: (i) to show that the smoothness of the potential plays no particular role in generating low-energy ABSs and (ii) to emphasize the difference between scenario ($\alpha$) -- which is often discussed in the context of a Gaussian potential barrier -- and scenario ($\beta$). Basically, the difference stems from the position of the chemical potential relative to the top of the potential hill. If $\mu \sim V_0^\prime$  and the potential hill is away from the end of the wire (i.e., it has nonzero occupancy on both sides), we are within the almost submerged potential hill scenario. By contrast, if $\mu \ll  V_0^\prime$ (with $\mu, V_0^\prime >0$) we have a (smooth) potential barrier that generates low-energy ABSs within scenario ($\alpha$) (see Sec. \ref{SecIIA}). In addition, we compare the low-energy ABS generated by a potential hill  ($V_0^\prime >0$) and the ABS induced by a potential well ($V_0^\prime <0$), both described by Eq. (\ref{Vx3}).

Figure \ref{FG5} shows the low-energy spectra of a system with an almost submerged potential hill defined by Eq. (\ref{Vx2}) -- panel (a) -- and  Eq. (\ref{Vx3}) -- panel (b), as well as  nearly dry potential well described by   Eq. (\ref{Vx3}) -- panel (c). 
The lowest-energy ABS mode localized near the potential hill/well (red line) collapses to zero energy at values of the Zeman field $\Gamma<\Gamma_c$ (i.e., in the topologically trivial regime).
The in-gap modes that collapse to zero energy at the TQPT ($\Gamma_c \approx 2~$meV) are associated with intrinsic Andreev bound states (i-ABSs) localized at the right end of the wire [see panels (a) and (b), blue in-gap lines], which emerge generically in systems with finite (positive) chemical potential and sharp confinement.\cite{Huang2018a} 
On the other hand, the low-energy in-gap modes in panels (b) and (c) with minima near $\Gamma \approx 0.9~$meV and $\Gamma \approx 1~$meV, respectively, correspond to additional ABSs localized at the potential hill/well (see the discussion of Figs.  \ref{FG7} and \ref{FG8} below). Finally, we note a significant difference between the ABS mode induced by a (smooth) potential hill [panel (b)], which is characterized by a very small energy splitting, and the ABS mode induced by a  potential well [panel (c)], which exhibits large energy splitting oscillations. Furthermore, unlike the oscillations characterizing overlapping MBSs in short wires, the amplitude of these oscillations {\em decreases} with increasing  Zeeman splitting.

In Fig. \ref{FG6} we show the Majorana components of a low-energy ABS induced by an almost submerged potential hill, with both MBSs belonging to the {\em same} spin-split sub-band, as demonstrated by the sign of the spin density in Fig. \ref{FG6}(c).
Note that each component MBS wave function  is characterized by  two ``exponential tails'' (with different characteristic length scale $\xi_M^{(1)} > \xi_M^{(2)}$), the dominant one pointing toward lower potential. Hence, in the case of potential hills, the (dominant) ``tails'' of the two  Majorana component wave functions point in opposite directions. This conclusion is strengthened by the example shown in Fig. \ref{FG7}(b). In both cases, the orientation of the (main) tails results in a strong suppression of the energy splitting. 
In general, the energy splitting oscillations of the lowest energy ABS mode [see Fig. \ref{FG5}(a)] are small, as long as $L^* = x_+^{(2)} -  x_+^{(1)} > \xi_M^{(2)}$, i.e., as long as the separation of the component MBSs is large compared to the characteristic length scale of the (minor) tail. Note that, by contrast, in the case of a potential well (see Fig. \ref{FG8}), the (major) exponential tails of the component MBSs point toward each other and overlap significantly (for a given Majorana separation $L^*$), generating significantly  larger energy splitting oscillations  [see Fig. \ref{FG5}(c)], which are comparable to those occurring in a short wire of length $L^*$. In this context, it is worth pointing out that the observation of energy splitting oscillations in finite wires (which was proposed as the ``smoking gun'' evidence for the presence of MZMs\cite{DSarma2012}) requires the presence of Majorana modes with exponential tails pointing toward each other, which, of course, is always the case in ideal (uniform) systems with hard confinement. Note that in systems with nonhomogeneous effective potential near the ends of the wire (e.g., quantum dots in the uncovered, barrier regions), the exponential tails pointing toward the opposite end of the system may be strongly suppressed, making the observation of  energy-splitting oscillations difficult. Therefore, ensuring hard enough confinement should be an important concern in this type of experiment. 

\begin{figure}[t]
\begin{center}
\includegraphics[width=0.49\textwidth]{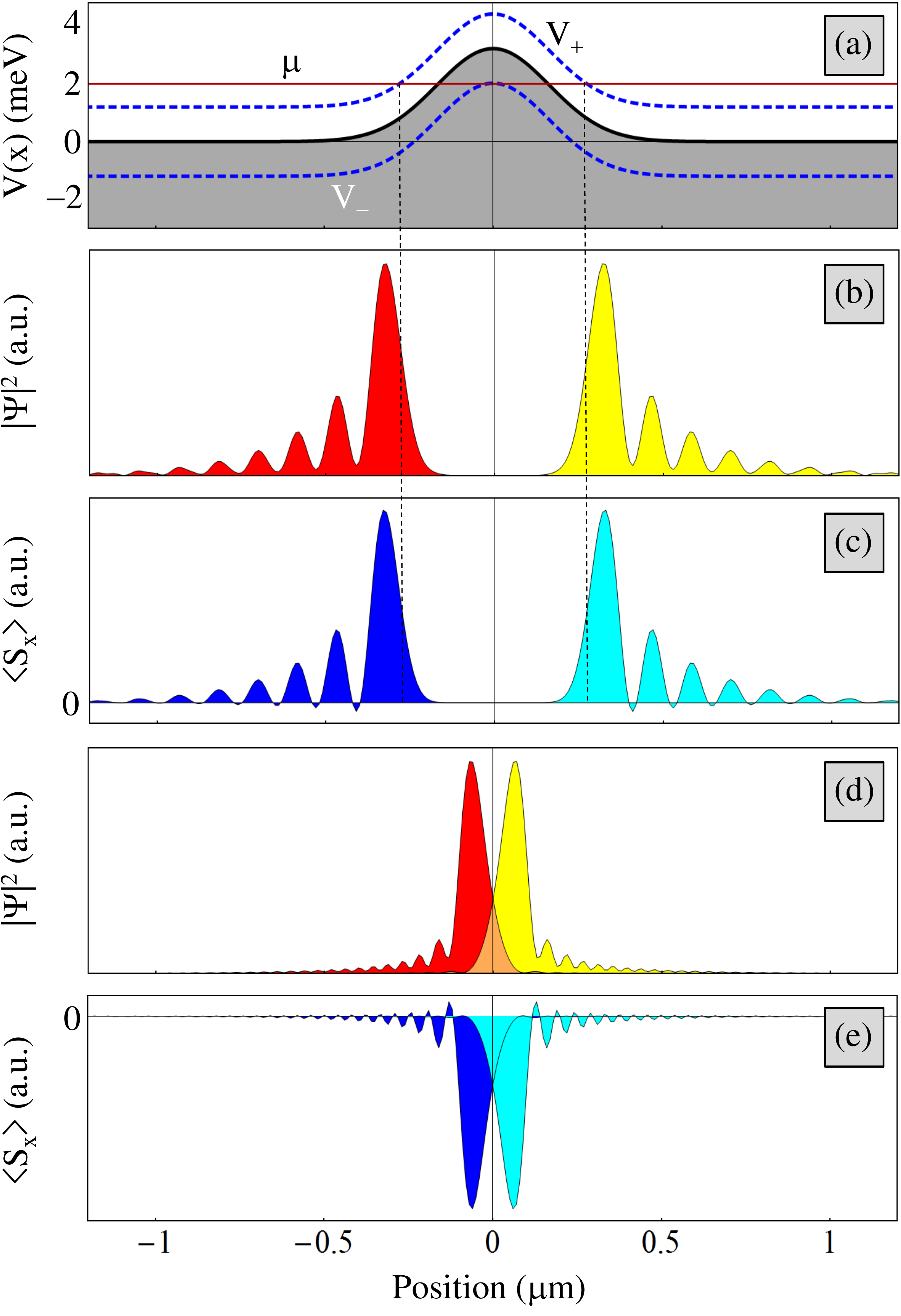}
\end{center}
\caption{(a) Potential profile corresponding to Eq. (\ref{Vx3}) with  $V_0^\prime=3.2~$meV and $\delta_{\rm v}=0.25~\mu$m (shaded black line). The dashed blue lines correspond to $V_\pm(x) = V(x) \pm\Gamma$, where $\Gamma=1.5~$meV is the Zeeman field. The chemical potential $\mu=2~$meV (red line) intersects the $V_+$ profile at $x_+^{(1)}\approx-0.25~\mu$m and $x_+^{(2)}\approx 0.25~\mu$m. (b) Majorana wave functions associated with the near-zero energy ABS marked `I' in Fig. \ref{FG5}(b). (c) Spin density of the Majorana modes shown in panel (b) revealing that two Majoranas are associated with the {\em same} spin-split sub-band.  (d) Majorana components of the ABS marked `II' in Fig. \ref{FG5}(b).  (e) Spin density of the Majorana modes shown in panel (d). Note that the separation of the two component MBSs is less than the the width of the main MBS peak. Consequently, the two MBSs have a significant overlap [orange area in (d)] and acquire a finite gap [see Fig. \ref{FG5}(b)].}
\label{FG7}
\end{figure}

The results shown in Fig. \ref{FG7} indicate that the key features of the low-energy ABS generated by an almost submerged potential hill (i.e., the component Majoranas are associated with the same spin-split sub-band and have exponential tails pointing away from each other)
do not depend on the smoothness of the potential. Importantly,  in this case the slightly raised top of the effective potential suppresses the secondary exponential tails that characterize the component MBS wave functions generated by a flat-top potential hill (see Fig. \ref{FG6}). Consequently, the overlap of the component MBSs (and the associated energy splitting) becomes negligible when $L^* > \delta_M$, i.e. when the separation of the two MBSs is larger than the width of the main MBS peak. 
While the ``spin-up'' sub-band generates two well-separated MBSs [see Fig. \ref{FG7}(b-c)] responsible for the near-zero-energy mode I in Fig. \ref{FG5}(b), the ``spin-down'' sub-band generates two \textit{overlapping} MBSs that give rise to a gapped ABS mode [see  mode II in Fig. \ref{FG5}(b)], as shown in  Fig. \ref{FG7}(d).
Note that in the regime $\mu < V_0^\prime - \Gamma$ the four Majorana modes shown in Fig. \ref{FG7} generate two type-($\alpha$) ABSs localized on the opposite slopes of the potential hill.  More generally, scenarios ($\alpha$) and ($\beta$) should be viewed as the {\em basic} single-band mechanisms for the emergence of low-energy ABSs in nonhomogeneous SM-SC hybrid structures. A generic low-energy ABS results from a combination of these basic mechanisms, with relative weights that depend on the details of the position-dependent effective potential and the value of the chemical potential.

\begin{figure}[t]
\begin{center}
\includegraphics[width=0.49\textwidth]{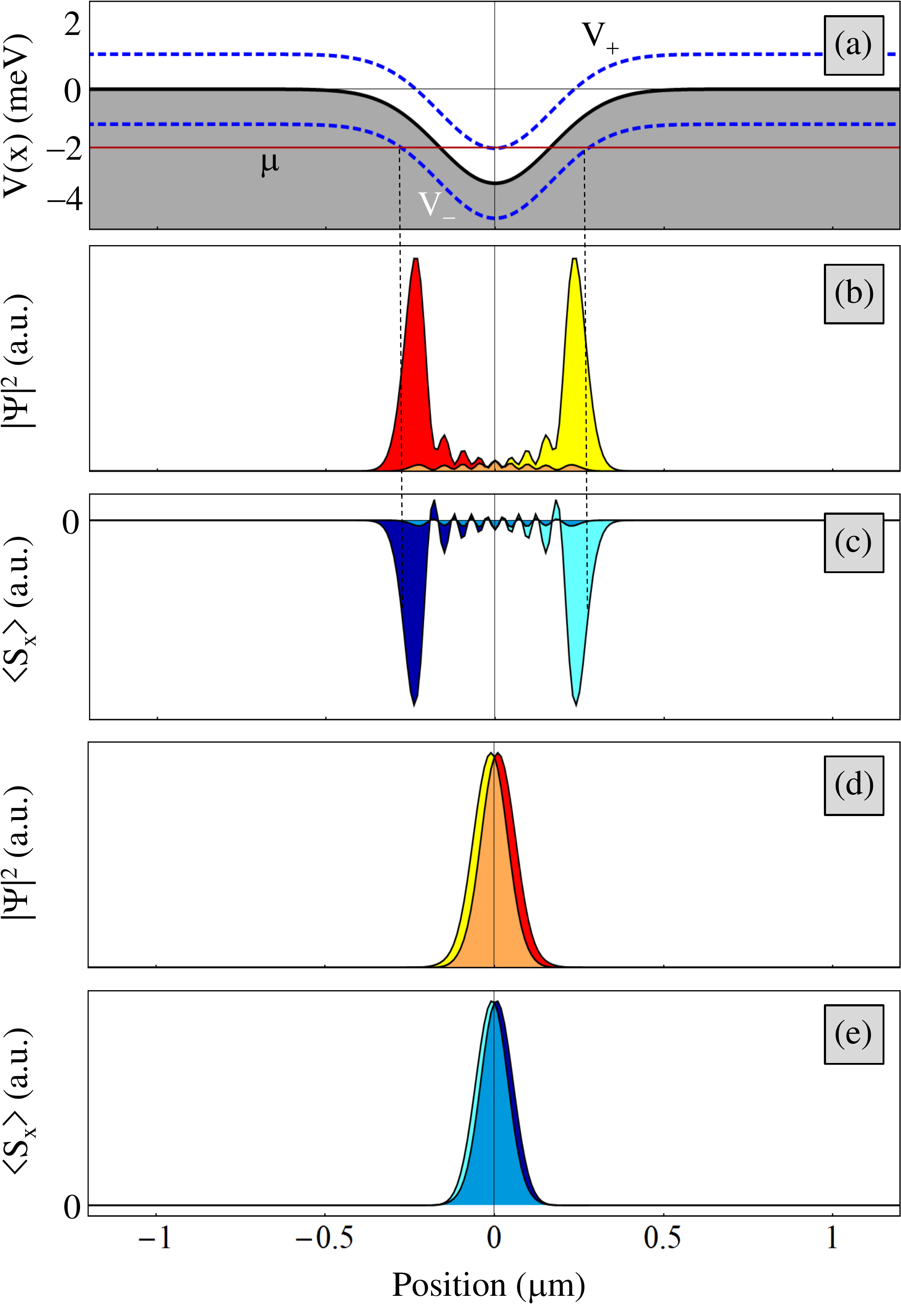}
\end{center}
\caption{(a) Potential profile corresponding to Eq. (\ref{Vx3}) with  $V_0^\prime=-3.2~$meV and $\delta_{\rm v}=0.25~\mu$m (shaded black line). The dashed blue lines correspond to $V_\pm(x) = V(x) \pm\Gamma$, where $\Gamma=1.5~$meV is the Zeeman field. The chemical potential $\mu=-2~$meV (red line) intersects the $V_-$ profile at $x_+^{(1)}\approx-0.25~\mu$m and $x_+^{(2)}\approx 0.25~\mu$m. (b) Majorana wave functions associated with the near-zero energy ABS marked I in Fig. \ref{FG5}(c). (c) Spin density of the Majorana modes shown in panel (b) revealing that two Majoranas are associated with the {\em same} spin-split sub-band.  (d) Majorana components of the ABS marked II in Fig. \ref{FG5}(c).  (e) Spin density of the Majorana modes shown in panel (d).}
\label{FG8}
\end{figure}

Finally,  in Fig. \ref{FG8} we show a ps-ABS generated by a potential well. The the ``spin-down'' sub-band generates two partially-separated MBSs [see panels (b-c)] responsible for the lowest-energy mode I in Fig. \ref{FG5}(c), while the ``spin-up'' sub-band generates two \textit{overlapping} MBSs  [see panels (d-e)] that give rise to a gapped ABS mode [see mode `II' in Fig. \ref{FG5}(c)].
The main difference, as compared to the potential hill case shown in Fig. \ref{FG7}, is the orientation of the exponential tails of the MBS wave functions toward each other. This results in the overlap of the tails and generates significant energy-splitting oscillations, as shown in Fig. \ref{FG5}(c). Note that key element here is the the orientation of the tails toward each other, rather than away from each other (which is the case for potential hills) or along the same direction (for ps-ABSs induced by finite width potential shores). In particular, the type-$\alpha$ ps-ABSs are characterized by relatively small energy splittings (see, e.g., Fig. \ref{FG1}), despite significant wave function overlap (see Figs. \ref{FG2}-\ref{FG3}). In Sec. \ref{SecIIIB2} we show explicitly that the suppression of the energy splitting associated with MBSs having the exponential tails pointing along the same direction [i.e.,  type-($\alpha$) ps-ABSs]  is not due to the MBSs having  opposite spin polarizations (see Fig. \ref{FG18}). Furthermore, in Sec. \ref{SecIIIB3} we show that the energy-splitting oscillations associated with MBSs having exponential tails pointing toward each other [i.e., type-($\beta$) ABSs induced by potential wells] are independent on the relative sign of the spin polarizations (see Figs. \ref{FG20}--\ref{FG22}).

We conclude this section with a brief summary of the basic physics responsible for the low-energy phenomenology in nonhomogeneous hybrid structures. In essence, in nonhomogeneous systems the “Majorana condition” can be satisfied locally within (using the language introduced above) shore potential regions, nearly dry potential well bottoms, and almost submerged potential hill tops. Consequently, partially overlapping Majorana bound states are generated near the ends of these finite (and typically short) ``topological regions''. A pair of such MBSs constitutes a (partially-separated) low-energy Andreev bound state (ps-ABS). In the subsequent sections, we show explicitly that the partial separation of the component MBSs is the key physical property responsible for the collapse of the ps-ABS energy toward zero, as well as for the difference in the coupling of the component MBSs to  external leads and, more generally, to local probes.

\begin{figure}[t]
\begin{center}
\includegraphics[width=0.49\textwidth]{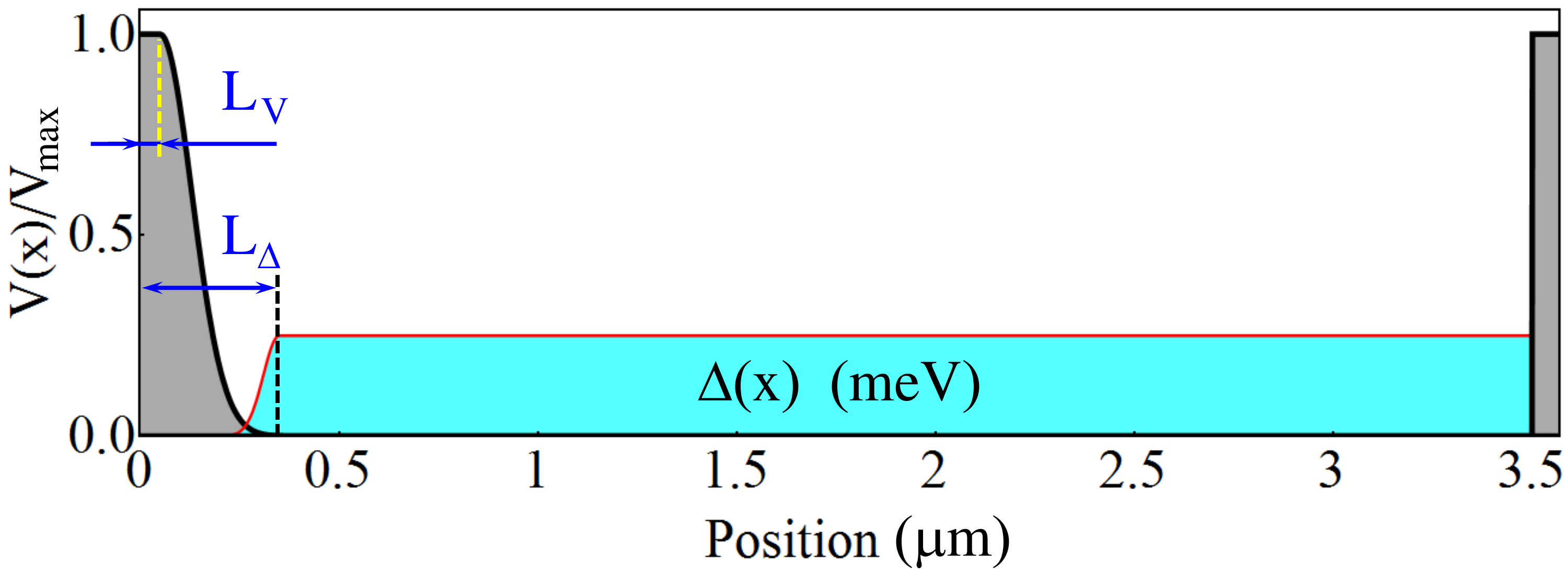}
\end{center}
\caption{Generic profile of the confining potential used in Sec. \ref{SecIII}  (gray-shaded line) and position-dependent proximity-induced pairing (cyan shading).}
\label{FG9}
\end{figure}

\section{Spatial separation and the collapse of Andreev modes to zero energy}\label{SecIII}

In Ref. [\onlinecite{vuik2018reproducing}] it was claimed that ABS states with near-zero energy can have fully overlapping MBS components, as long as  these states have an approximately opposite spin.
By contrast, in this section we show that obtaining a robust near-zero energy ABS mode in the topologically trivial regime of SM-SC heterostructures requires that the component MBSs be partially separated in real space. If we focus on ABSs generated by finite width potential shores, the component ABSs have to be separated by a distance larger than $\delta_M$, the characteristic width of the main peak of the Majorana wave function. Note that, within this scenario, the exponential tails of the Majorana wave functions point in the same direction and, consequently, the two MBSs can have a substantial overlap. Also note that we do not disagree with the technical results presented in Ref. [\onlinecite{vuik2018reproducing}], but point out the key role of the spatial MBS separation in generating the collapse of topologically-trivial ABS modes toward zero energy [e.g., the condition $L^* > \delta_M$ for type-($\alpha$) ps-ABSs].

Based on the analysis of the MBS wave functions, we also argue that the spatial separation of the component MBSs, together with details regarding the wave function profiles, are key generic factors that determine the signature of a low-energy ABS in an experiment based on local probes (e.g., charge tunneling into the end of the wire). By contrast, the association of the component MBSs with different spin-split sub-bands, or the fact that their spectral structure may involve different momentum components (i.e., different Fermi momenta) are rather specific properties (particularly relevant for a system with smooth confinement) that should not be seen as the ultimate ``cause'' of the collapse to zero energy of the ABS mode or of the local probes coupling very differently to the component MBSs. Below, we discuss a few examples that support this picture. 
Based on our analysis, we conclude that the low-energy ABS modes induced by nonhomogeneous potentials can be viewed most naturally as partially separated Andreev bound states (ps-ABSs).

The generic profile of the confining potential used in the numerical calculations discussed below is shown in Fig. \ref{FG9}. The left end of the wire has soft confinement given by
\begin{equation}
V(x) =V_{max}\times \left\{
\begin{array}{l}
1  ~~~~~~~~~~~~~~~~~~~~~\!~~~ {\rm if~}~ x< L_{\rm v}, \\
\exp\left[-\frac{(x-L_{\rm v})^2}{\delta_{\rm v}^2}\right] ~~~~ {\rm if~}~  L_{\rm v}<x<L, \\
1  ~~~~~~~~~~~~~~~~~~~~~\!~~~ {\rm if~}~ x> L,
\end{array}\right.    \label{VV1}
\end{equation}
where $L_{\rm v}$ defines a ``flat top'' region and $\delta_{\rm v}$ is a parameter that controls the smoothness of the potential, with $\delta_{\rm v}\rightarrow 0$ corresponding to the hard wall limit. We also consider the  situation in which the end of the wire (i.e., the barrier region) is not proximitized by having a position-dependent induced pair potential, as shown in Fig. \ref{FG9}. Unless specified otherwise, we will take $L_{\rm v}=0$, $L_\Delta=0$, and $L=3.5~\mu$m.

\begin{figure}[t]
\begin{center}
\includegraphics[width=0.49\textwidth]{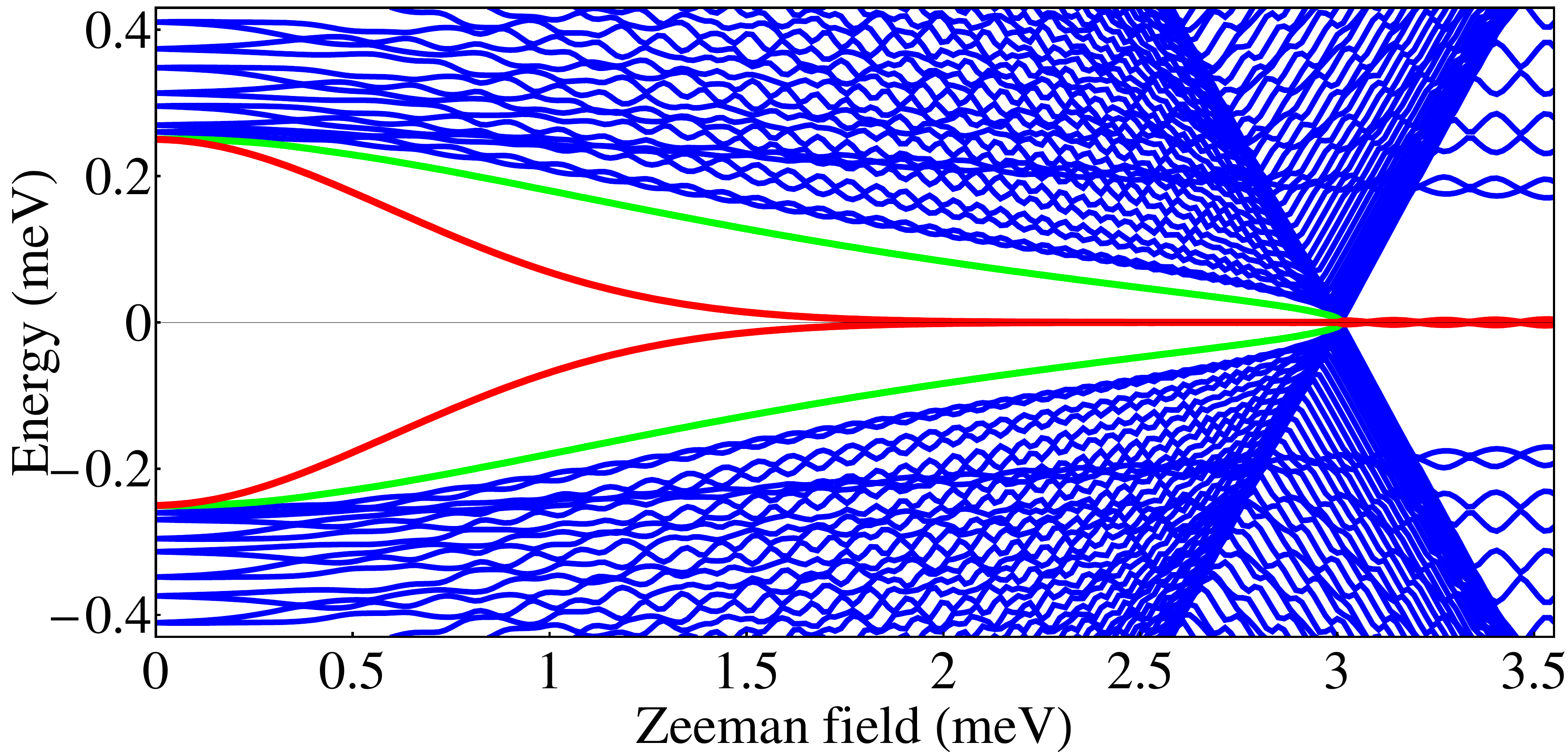}
\end{center}
\caption{Dependence of the low-energy spectrum on the applied Zeeman field for a system with effective potential given by Eq. (\ref{VV1}) and chemical potential $\mu=3~$meV. The potential parameters are $V_{max}=7~$meV and $\delta_{\rm v}=0.1~\mu$m. The green lines correspond to the i-ABS localized at the right end of the wire, while the lowest energy mode (red lines) is associated with the ps-ABS localized at the left end (for $\Gamma < 3~$meV) or the MZMs localized at the opposite ends of the system (above the TQPT).}
\label{FG10}
\end{figure}

\subsection{From i-ABSs to ps-ABSs in systems with smooth confinement}\label{SecIIIA}

In this section we establish that the near-zero energy ABS modes emerging in the trivial phase in systems with smooth confinement are adiabatically connected to the intrinsic ABSs that are generically present in a clean, homogeneous  wire (with hard confinement at the ends). At finite Zeeman field, the smooth confinement partially separates the component MBSs of the Andreev mode, which results in its collapse to zero energy.

\begin{figure}[t]
\begin{center}
\includegraphics[width=0.49\textwidth]{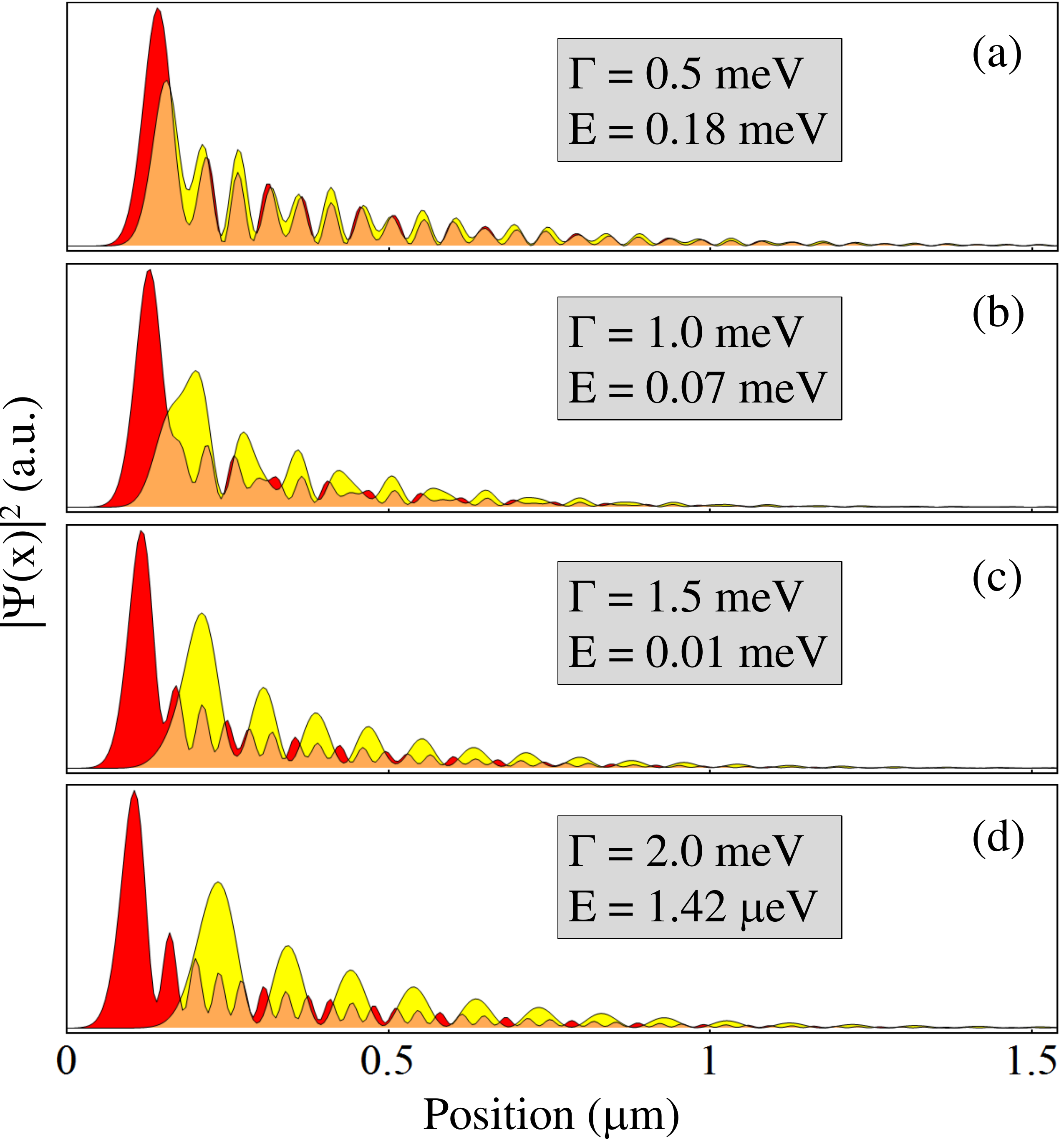}
\end{center}
\caption{Majorana wave functions [given by Eqs. (\ref{chiA} and \ref{chiB})] associated with the lowest-energy  ABS mode corresponding to the red line in Fig. \ref{FG10} for different values of the Zeeman field $\Gamma < \Gamma_c=3~$meV. Note that the ABS energy decreases with increasing separation between the component MBSs and approaches zero (i.e., $E \ll \Delta=0.25~$meV) when the separation is larger that the characteristic width of the main  peak of the Majorana wave function.}
\label{FG11}
\end{figure}

Consider a clean wire with finite chemical potential. Generically, it supports sub-gap i-ABSs localized at the ends of the wire\cite{Huang2018a} (green line in Fig. \ref{FG10}). Upon softening the confinement at one end of the wire (see Fig. \ref{FG9}), the corresponding sub-gap mode collapses toward zero-energy (red line in Fig. \ref{FG10}).
In Fig. \ref{FG11} we show that the collapse toward zero energy of the ABS mode localized at the left end of the wire is correlated with a Zeeman field-dependent spatial separation of the component MBSs.
The positions of the main peaks of the MBS wave functions are determined by the solutions $x_\pm$ of Eq. (\ref{Vpm}) and the corresponding separation is $L^*(\Gamma) = x_+(\Gamma)-x_-(\Gamma)$. The condition for well-separated component MBSs [within scenario ($\alpha$)] is $L^* > \delta_M$, i.e., the separation should exceed the characteristic width of the main peak of the Majorana wave function.
Note that within this scenario two well-separated component MBSs (whose ``exponential tails'' point in the same direction) can still have a substantial overlap (orange areas in Fig. \ref{FG11}). This makes them extremely susceptible to local perturbations (hence completely unprotected). By contrast,  a ps-ABS generated by an almost submerged potential hill (see, for example, Fig. \ref{FG7}), is significantly better protected against local perturbations (for any comparable value of the separation $L^*$).

\begin{figure}[t]
\begin{center}
\includegraphics[width=0.49\textwidth]{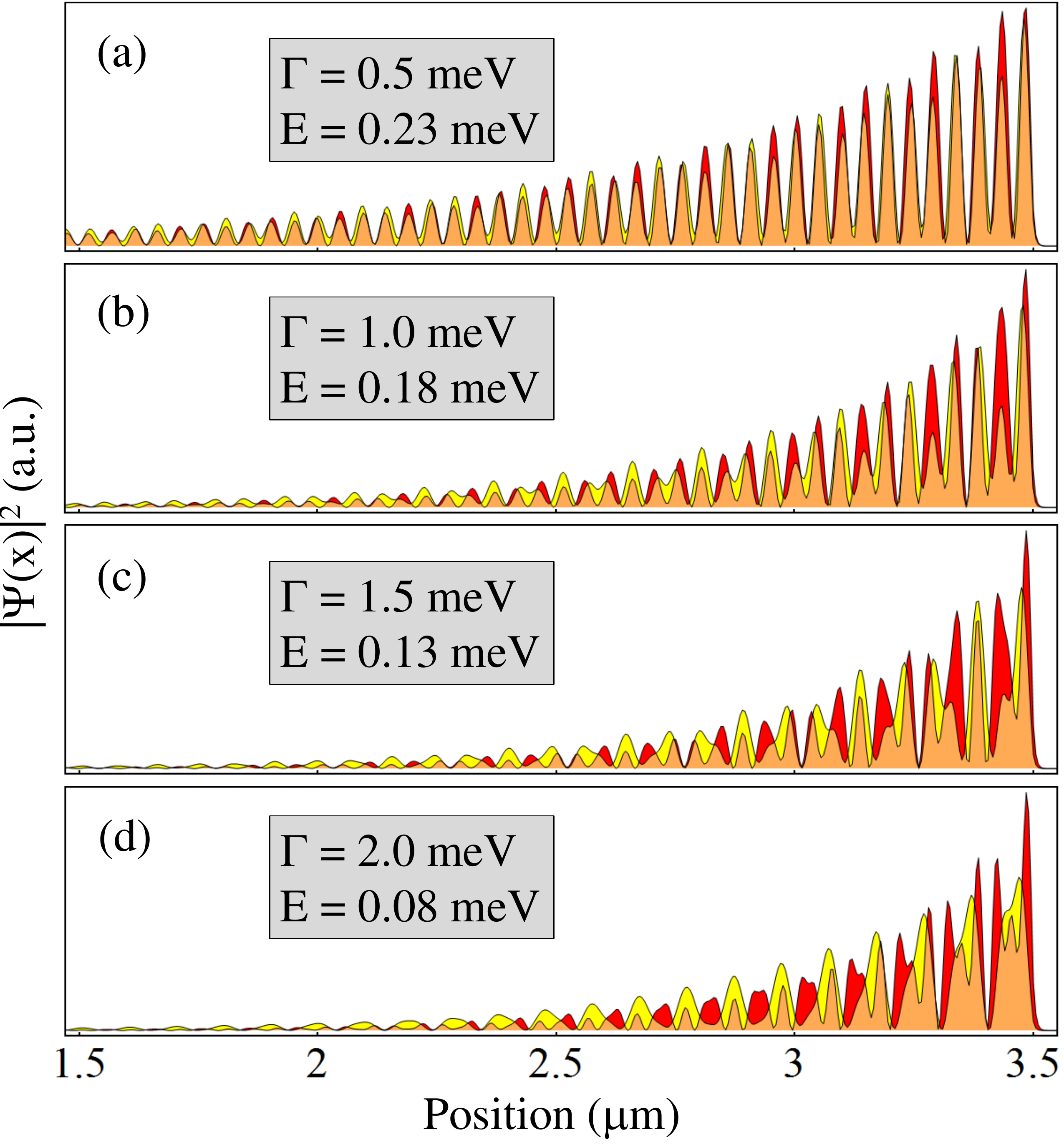}
\end{center}
\caption{Majorana wave functions associated with the i-ABS mode corresponding to the green line in Fig. \ref{FG10} for different values of the Zeeman field. Note that the component MBSs do not separate and the energy remains comparable to the bulk quasiparticle gap.}
\label{FG12}
\end{figure}

By contrast, in Fig. \ref{FG12} we show that the component MBSs of a sub-gap i-ABS are not spatially separated (as described above), which results in the i-ABS having a non-zero energy comparable to the bulk quasiparticle gap. Note, however, that the two component Majorana wave functions are not identical when $\Gamma >0$,  because they have different spectral composition (i.e., different characteristic wave vectors).
Also note that the confining potential at the right end of the wire is a finite step function (see Fig. \ref{FG9}), rather than an infinite wall, which allows the wave functions to partially penetrate into the barrier region. This penetration is different for the two component MBSs, being deeper for the MBS with larger characteristic wave vector (i.e., the ``red'' Majorana). Consequently, the component MBSs could, in principle, couple very differently to a local probe, similar to the MBSs at the left end of the wire. However, in the presence of a sharp potential (more precisely, a potential characterized by a vanishing ``shore width'') the energy of the in-gap ABS remains comparable to the bulk quasiparticle gap, being unable to produce a near-zero-bias conductance peak in local charge tunneling experiments.

\begin{figure}[t]
\begin{center}
\includegraphics[width=0.49\textwidth]{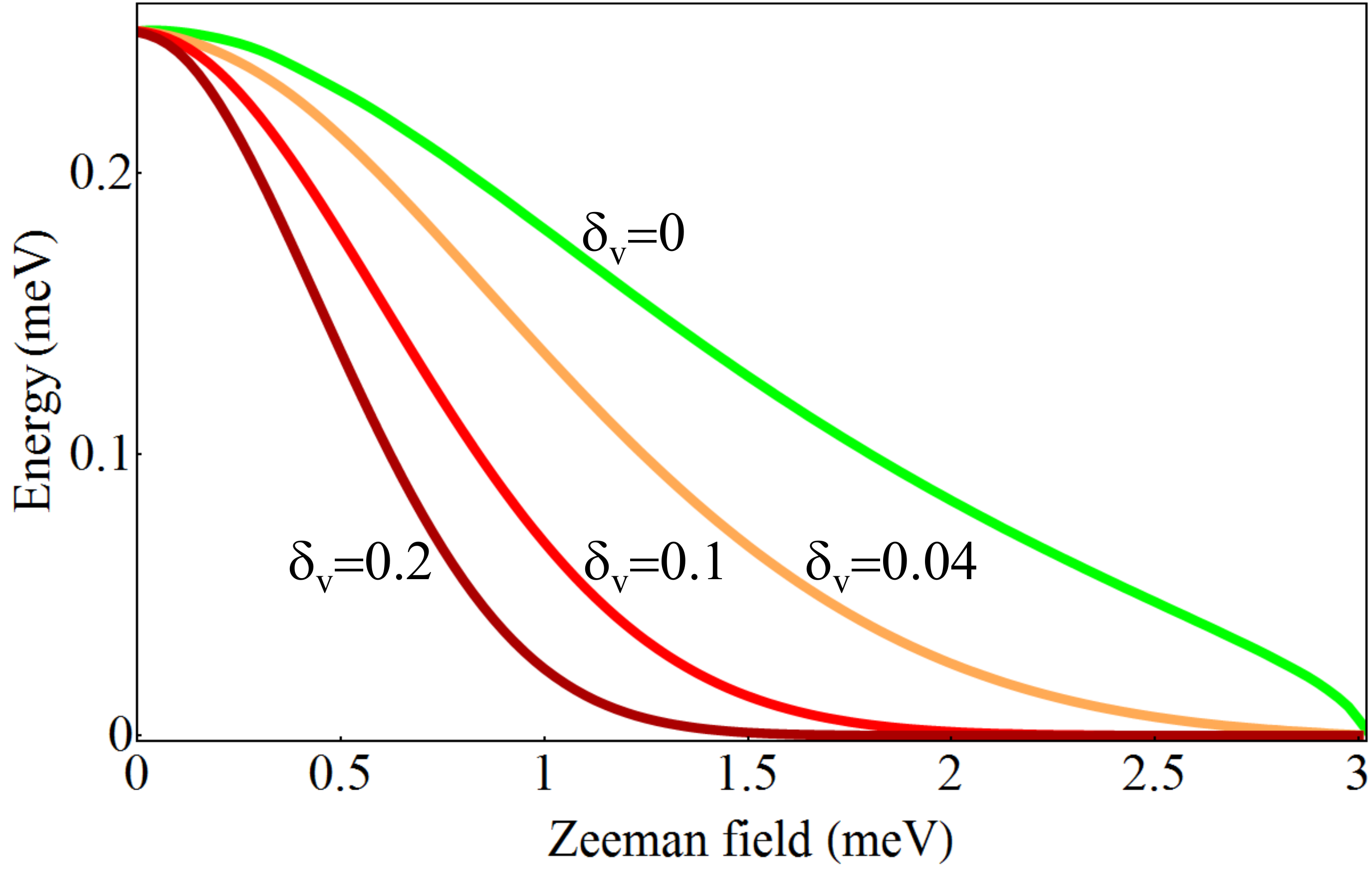}
\end{center}
\caption{The collapse to zero energy of the i-ABS mode ($\delta_{\rm v}=0$) in the presence of smooth confinement ($\delta_{\rm v}\neq 0$). The green line coincides with the corresponding mode in Fig. \ref{FG10}. The parameter $\delta_{\rm v}$ describing the smoothness of the confining potential is given in microns.}
\label{FG13}
\end{figure}

The adiabatic connection between the ps-ABS mode that emerges in the presence of smooth confinement and the i-ABS occurring generically in clean proximitized wires with finite chemical potential is illustrated in Fig. \ref{FG13}. 
Note that, for a given value of the Zeeman field, the separation of the component MBSs increases monotonically with $\delta_{\rm v}$ (i.e., with decreasing the average slope). Hence, the crossover field $\Gamma_c^*$ associated with the collapse to zero energy of the ps-ABS mode decreases with $\delta_{\rm v}$.

\subsection{The role of spin}\label{SecIIIB}

In this section we show that, in general, the spin structure of the component MBSs does not play a fundamental role in i) generating vastly different coupling strengths of the component MBSs to local probes and ii) ensuring the robustness of the near-zero-energy ABS mode. The common picture regarding the role of spin is based on the implicit assumption that the low-energy ABS is generated within scenario ($\alpha$) by the tunnel barrier itself. However, explicit position-dependent Schr\"{o}dinger-Poisson calculations have demonstrated\cite{Woods2018} that the potential profile in a proximitized nanowire can have inhomogeneities generated by other sources and can be located away from the barrier region. Below, we show that, in general, it is the spatial profile of the Majorana wave functions (rather that its spin structure) that determines the properties of a low-energy ABS.

\subsubsection{Asymmetric channel potential}\label{SecIIIB1}

\begin{figure}[t]
\begin{center}
\includegraphics[width=0.49\textwidth]{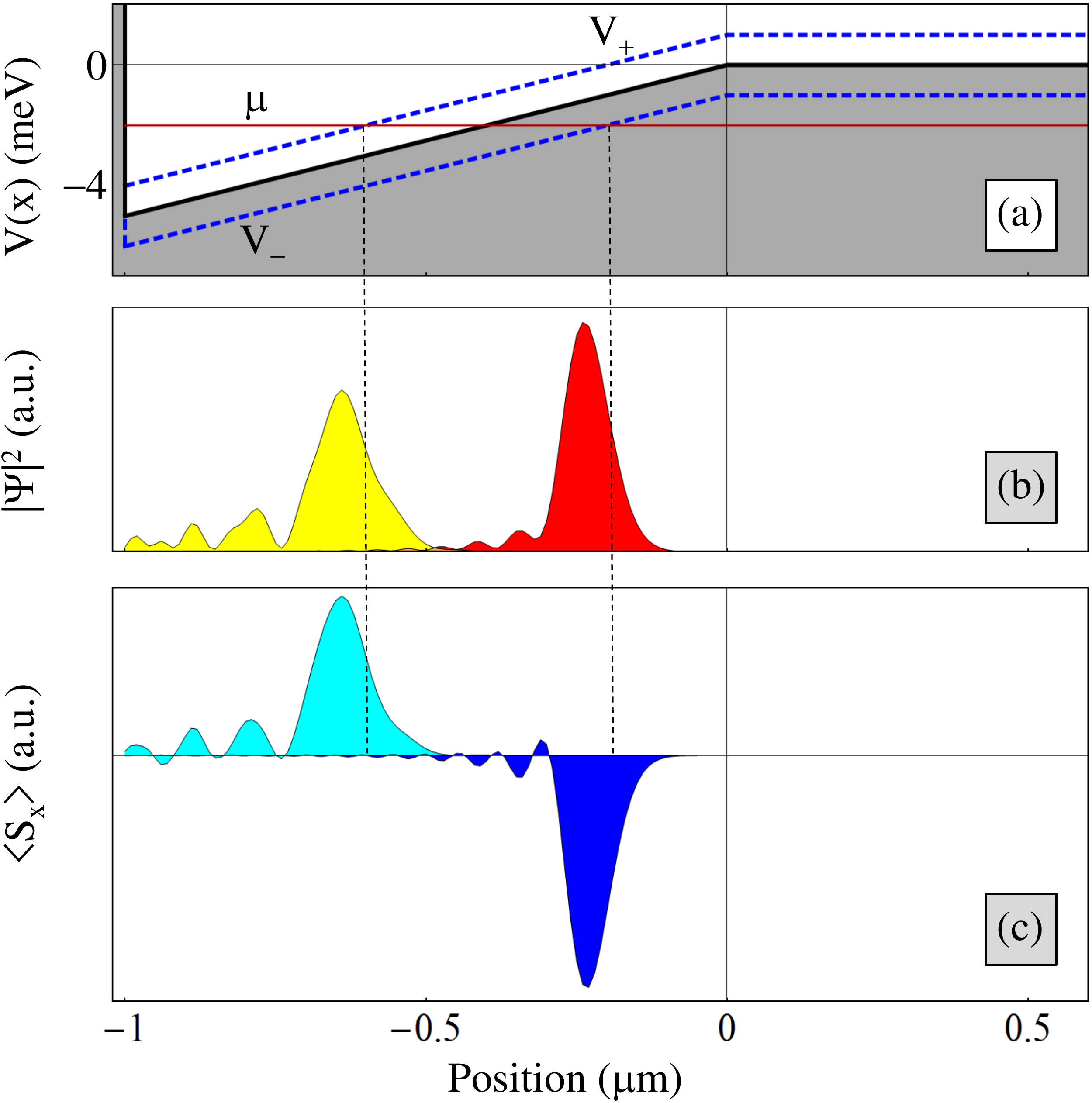}
\end{center}
\caption{(a) Asymmetric channel potential profile (shaded black line) and the associated $V_\pm$ profiles (dashed blue lines) corresponding to $\Gamma=1~$meV.  (b) Majorana wave functions $\psi_A$ (red) and $\psi_B$ (yellow) [given by Eqs. (\ref{chiA}-\ref{chiB})] associated with the near-zero energy ABS marked by a small circle in Fig. \ref{FG15}. (c) Spin density of the Majorana modes shown in (b).}
\label{FG14}
\end{figure}

We consider the asymmetric ``channel'' potential shown in Fig. \ref{FG14}(a) consisting of a sharp barrier (of unspecified height) at the left end of the wire and a constant positive slope within a $1~\mu$m segment. This potential profile effectively defines a quantum dot at the left end of the wire. At finite Zeeman field, a ps-ABS is generated within scenario ($\alpha$), having spatially-separated component MBSs associated with opposite spin-split sub-bands, as shown in see Fig. \ref{FG14}, panels (b) and (c).

\begin{figure}[t]
\begin{center}
\includegraphics[width=0.49\textwidth]{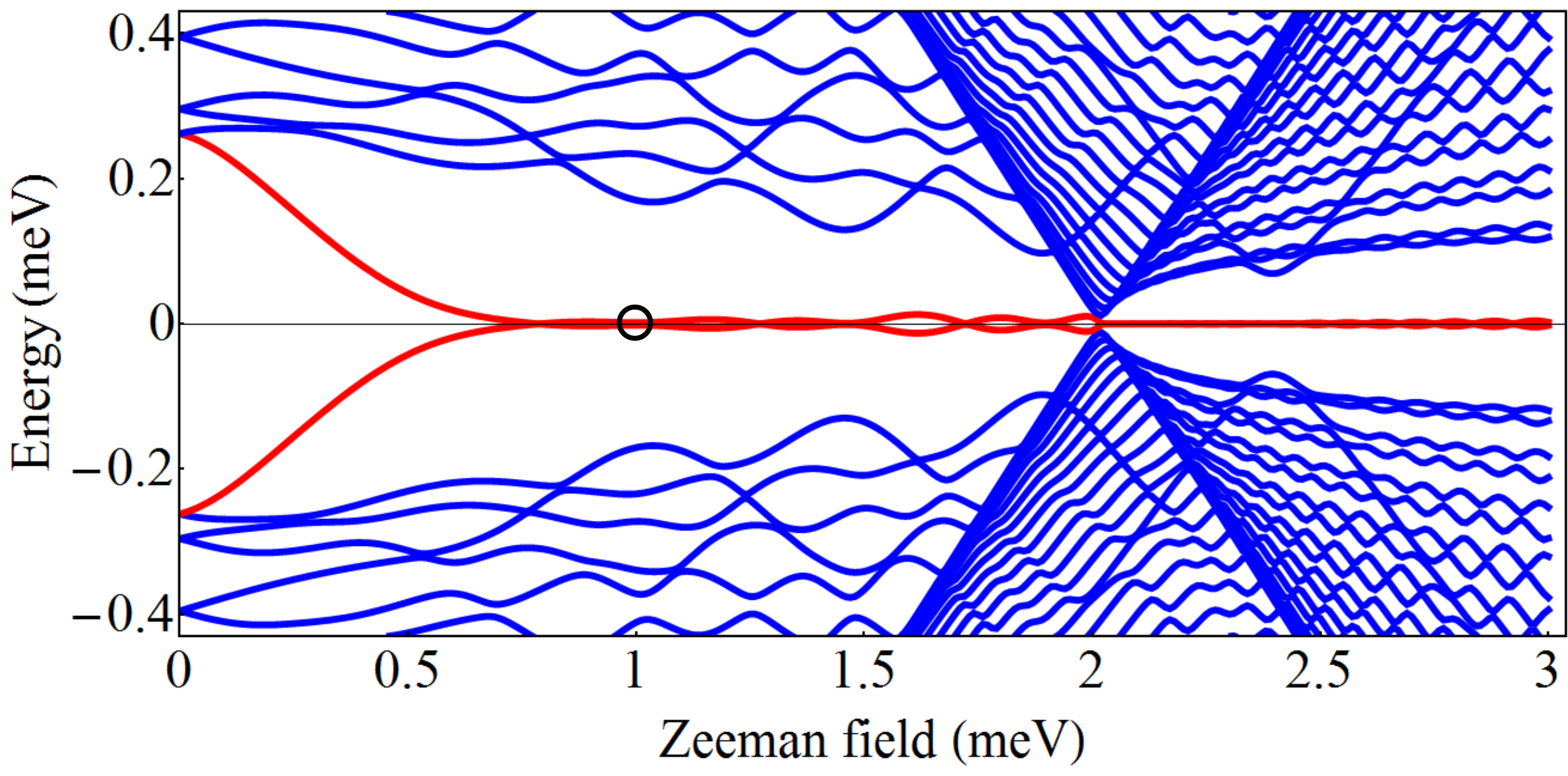}
\end{center}
\caption{Dependence of the low-energy spectrum on the applied Zeeman field for a system with a quantum dot defined by the asymmetric channel potential shown in Fig. \ref{FG14}(a) and chemical potential $\mu=-2~$meV. The wave functions of the component MBSs of the state marked by the small circle are shown in Fig. \ref{FG14}.}
\label{FG15}
\end{figure}

The low-energy spectrum of a system with asymmetric ``channel'' potential [as shown in Fig. \ref{FG14}(a)] is given in Fig. \ref{FG15}. Note that the critical field associated with the TQPT is $\Gamma_c=|\mu|=2~$meV. A near-zero energy ABS mode  induced by the position-dependent potential emerges at the left end of the wire for $\Gamma <\Gamma_c$. The state marked by a small circle is shown in Fig. \ref{FG14}.
While in the ``standard'' case (see, for example, Figs. \ref{FG9}-\ref{FG11})  the spin-down (i.e., `red') Majorana couples strongly to a local probe placed at the left end of the wire and the spin-up (i.e., `yellow') Majorana couples weakly, it is important to note that in this situation the coupling strengths are reversed. The typical explanation for the different coupling strengths corresponding to the two modes  is that the two MBSs experience different effective barriers (with the `red' Majorana experiencing a lower barrier). While the height of the effective barrier would always be relevant for propagating modes, the MBSs are localized modes and the decisive factor that controls their coupling to a local probe is the spatial profile of the wave function, as can be easily seen by comparing Figs. \ref{FG11} and \ref{FG14}. We emphasize that in both situations the $V_-$ left barrier is lower than the $V_+$ barrier. Note also that, in contrast with the ``standard'' smooth-confinement situation,\cite{vuik2018reproducing}  completely  suppressing the  left (sharp) barrier does not destroy the ps-ABS. Furthermore, a left probe will still couple strongly to the `yellow' (spin-up) Majorana and weakly to the `red' (spin-down) MBS.

\subsubsection{Low-energy modes in two-band systems}\label{SecIIIB2}

\begin{figure}[t]
\begin{center}
\includegraphics[width=0.49\textwidth]{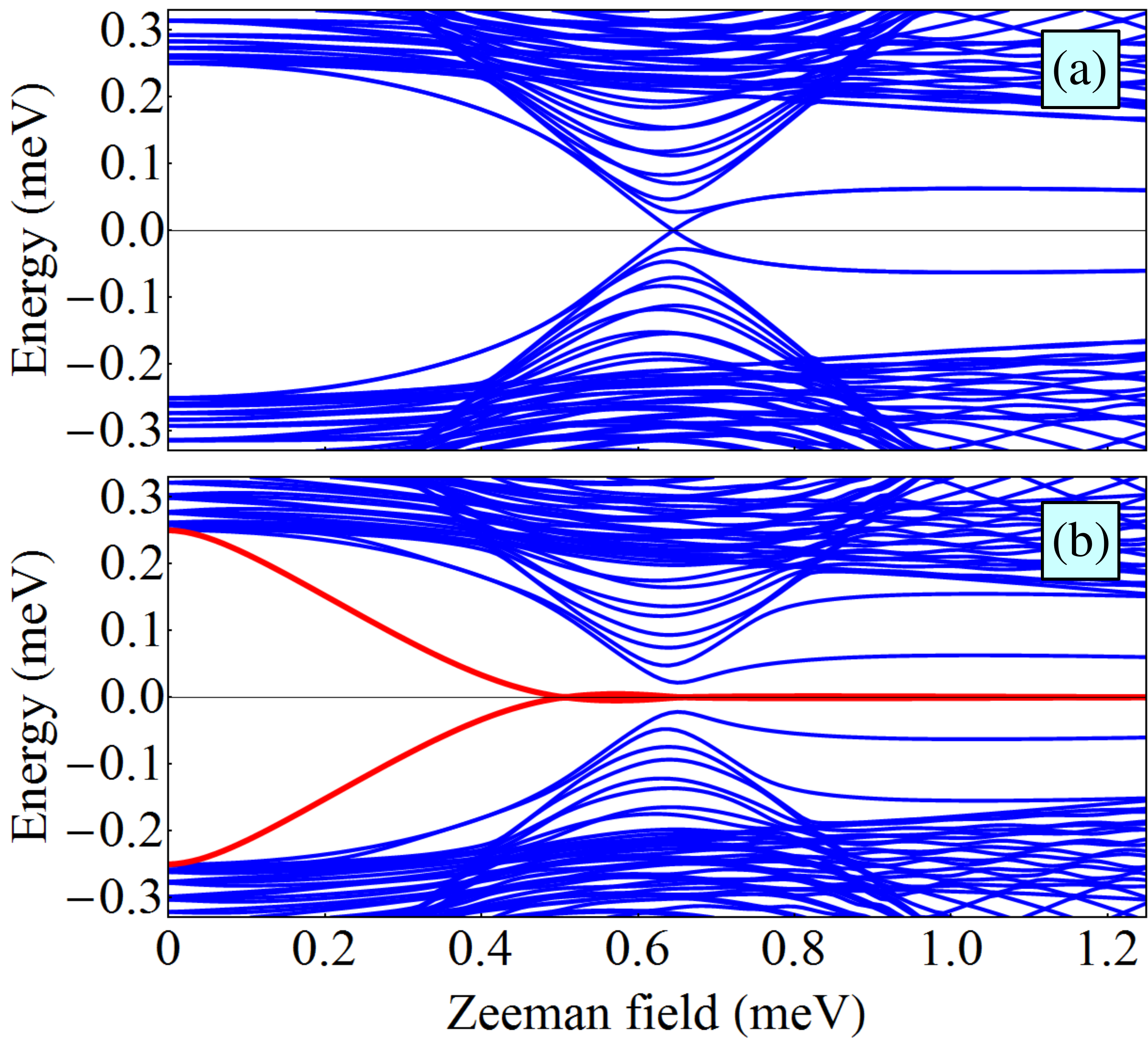}
\end{center}
\caption{Dependence of the low-energy spectrum on the applied Zeeman field for a two-band system with (a) hard confinement and (b) soft confinement at the left end of the wire. Note that the soft confinement induces a near-zero-energy mode [red line in panel (b)] that extends on both sides of the crossover point $\Gamma^*\approx 0.65~$meV. The nature of different sub-gap modes becomes clear by examining the corresponding wave functions (see Fig. \ref{FG18}). Model parameters: inter-band gap $\Delta\epsilon = 0.6~$meV, chemical potential $\mu=0.3~$meV, longitudinal spin-orbit coupling $\alpha=1~$meV, transverse spin-orbit coupling $\alpha^\prime=0.5~$meV, barrier smoothness $\delta_{\rm v}=0.25~\mu$m, and barrier height $V_{\rm max}=2~$meV [see Eq. (\ref{VV1})].}
\label{FG16}
\end{figure}

\begin{figure}[t]
\begin{center}
\includegraphics[width=0.49\textwidth]{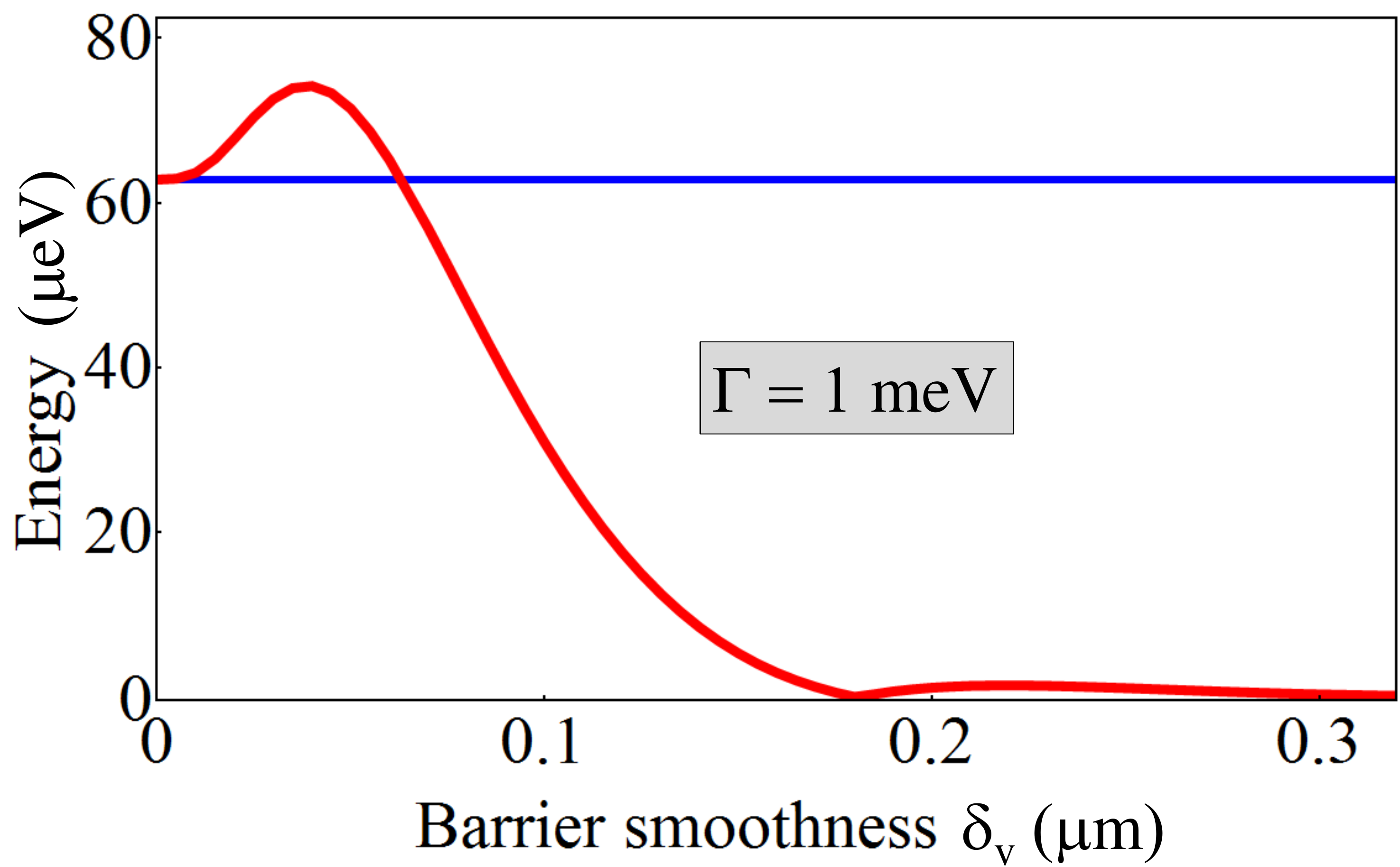}
\end{center}
\caption{The collapse toward zero-energy of the ABS mode localized near the left end of the wire as function of the barrier smoothness $\delta_{\rm v}$ in a two-band system with the same parameters as in Fig. \ref{FG16}. The blue line gives the energy of the ABS in a system with hard confinement, $\delta_{\rm v}=0$ (provided for comparison).}
\label{FG17}
\end{figure}

In this section we study a two-band model by considering two ``copies'' of the single-band  Hamiltonian given by Eq. (\ref{H}). The two bands are separated by an inter-band gap $\Delta\epsilon = 0.6~$meV and are coupled by a transverse Rashba term $H_{12} = i \frac{\alpha^\prime}{2}\sum_j(c_j^\dagger \sigma_x c_j + h.c.)$.  We tune the chemical potential midway between the two bands, $\mu =0.3~$meV. In the absence of a transverse Rashba term ($\alpha^\prime = 0$), the system is characterized by a TQPT separating a gapped topologically trivial phase from a topological phase characterized by the presence of two Majorana modes at each end of the wire. These modes are protected by an additional chiral symmetry \cite{Sumanta1, Sumanta2} and are associated with the spin-down sub-bands. In the presence of a finite transverse Rashba term, this in-gap mode acquires a finite gap, as shown in Fig. \ref{FG16}.
In panel (a) we show the low-energy spectrum of a (coupled) two-band system with hard confinement, while panel (b) corresponds to a system with soft confinement at one of the ends [given by Eq. (\ref{VV1}) with $\delta_{\rm v}=0.25~\mu$m]. Note that the wire with hard confinement at both ends is gapped both below and above the crossover field $\Gamma^*\approx 0.65~$meV corresponding to the minimum of the bulk quasiparticle gap. By contrast, the soft confinement induces the collapse of the ABS mode to zero energy [red line in panel (b)].
The effect of soft confinement on the in-gap ABS mode is illustrated in Fig. \ref{FG17} for a fixed value of the Zeeman field. Note that the energy of the ABS mode collapses toward zero for $\delta_{\rm v} > 0.18~\mu$m.

\begin{figure}[t]
\begin{center}
\includegraphics[width=0.49\textwidth]{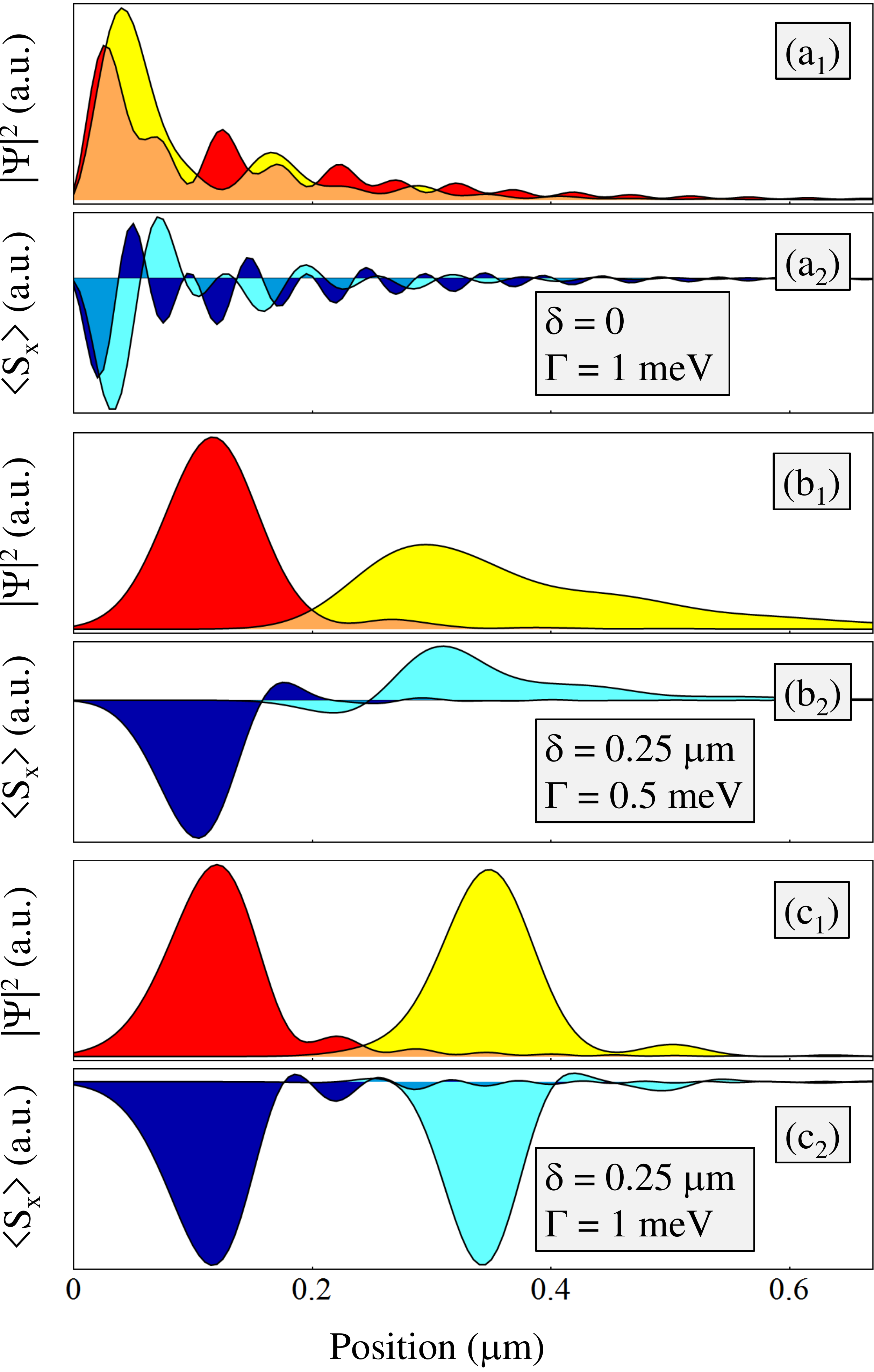}
\end{center}
\caption{Majorana wave functions and the corresponding spin densities for different in-gap ABS modes  characterizing the two-band system from Fig. \ref{FG16}. (a) ABS mode localized at the left end of the wire in a system with hard confinement ($\delta_{\rm v}=0$) and Zeeman field $\Gamma=1~$meV [see Fig. \ref{FG16}(a)]. Note that the two MBSs are not separated and consist of spin-up and spin-down contributions. (b) ABS mode localized at the left end of the wire in a system with soft confinement ($\delta_{\rm v}=0.25~\mu$m) and Zeeman field $\Gamma <\Gamma_c^*$ [see Fig. \ref{FG16}(a), red line].   (c) ABS mode localized at the left end of the wire in a system with soft confinement ($\delta_{\rm v}=0.25~\mu$m) and Zeeman field $\Gamma >\Gamma_c^*$ [see Fig. \ref{FG16}(a), red line].}
\label{FG18}
\end{figure}

To clarify the physics behind the emergence of the near-zero-energy mode, we calculate the Majorana wave functions for the component MBSs associated with different in-gap modes shown in Fig. \ref{FG16}, as well as the corresponding spin densities. The results are given in Fig. \ref{FG18}.
In panel (a) we show the Majorana wave functions of the finite energy ABS modes emerging in a two-band system with hard confinement and Zeeman field $\Gamma > \Gamma_c$ [see the in-gap modes from Fig. \ref{FG16}(a)]. Note that (i) the component MBSs are not separated spatially and (ii) they consist of a mixture of spin-up and spin-down contributions. A similar ABS is localized at the right end of the wire (not shown).
In panel (b) we show the Majorana wave functions of the near-zero energy ABS mode emerging in a two-band system with soft confinement and Zeeman field $\Gamma < \Gamma_c$ [see the red lines in Fig. \ref{FG16}(b)]. Note that the component MBSs have the same structure as those generated in a single-band system under scenario ($\alpha$). In particular, the two MBSs are spatially separated and are (predominantly) associated with  {\em different} spin-split sub-bands.
Finally, in panel (c) we show the Majorana wave functions of the near-zero energy ABS mode emerging in a two-band system with soft confinement and Zeeman field $\Gamma > \Gamma_c$ [see the red lines in Fig. \ref{FG16}(b)]. The two component MBSs are well separated, but, in contrast to panel (b), both are associated with spin-down sub-bands.

\begin{figure}[t]
\begin{center}
\includegraphics[width=0.49\textwidth]{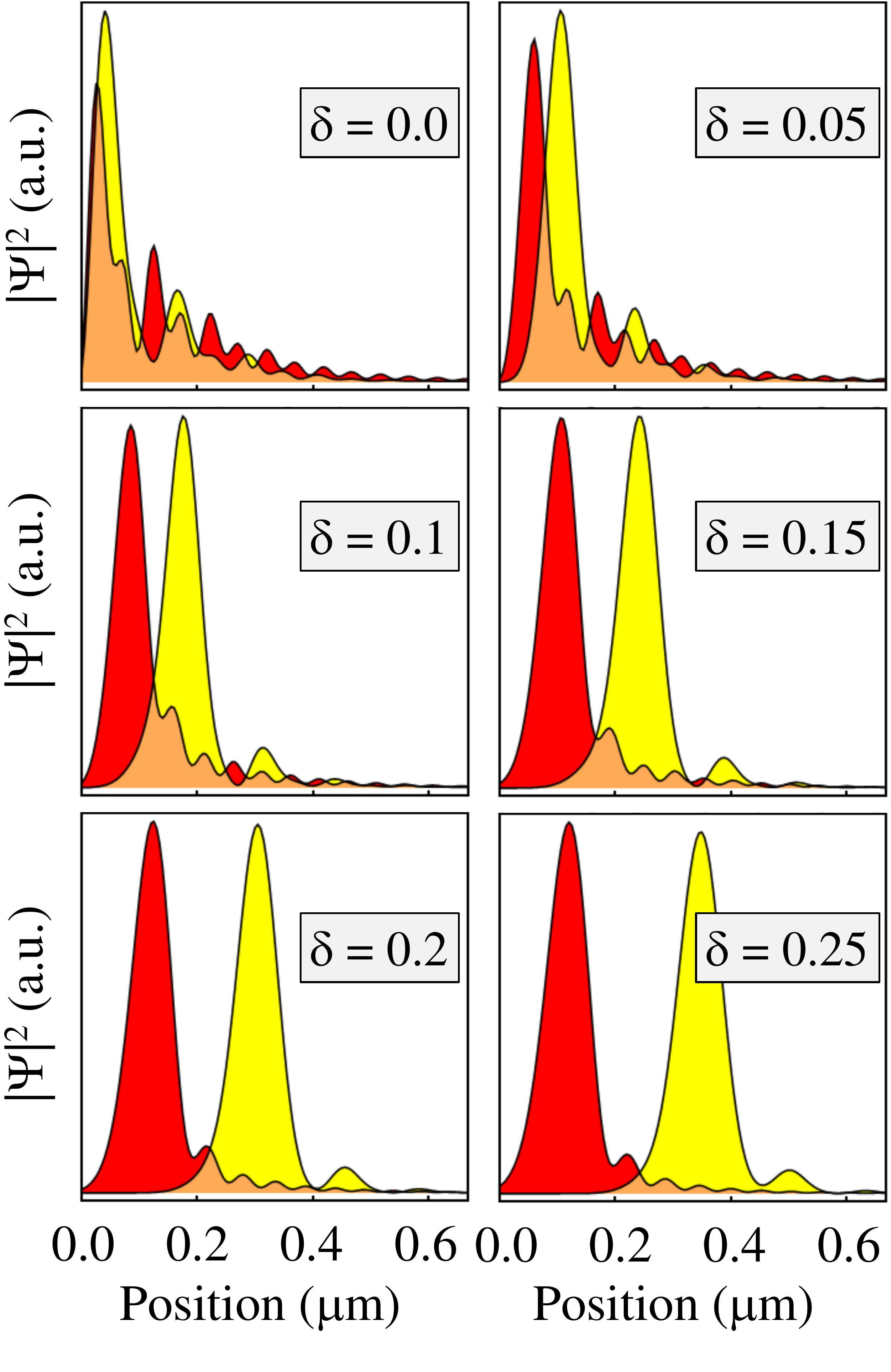}
\end{center}
\caption{Evolution of the Majorana wave function associated with the lowest-energy mode of a two-band system with smooth confinement and Zeeman field $\Gamma >\Gamma_c^*$. The corresponding energies are given in Fig. \ref{FG17}. Note that the ABS mode collapses toward zero energy when the separation of the component MBSs exceeds the characteristic width of the main Majorana peak.}
\label{FG19}
\end{figure}

The key role of the spatial separation of the component MBSs in the collapse to zero energy of the ABS mode in the regime $\Gamma > \Gamma_c$ (i.e., when the MBSs belong to the same type of spin-split sub-band) is revealed by the dependence of the wave functions on the barrier smoothness  shown in Fig. \ref{FG19}.  Similar to scenario ($\alpha$), which is valid within the single-band approximation, the low-energy ABS collapses to zero energy when the separation of the component MBSs exceeds the characteristic width of the main Majorana peak.
This example explicitly demonstrates that the spin structure of the component MBSs is irrelevant in determining the collapse to zero-energy of the low-energy ABS mode. By contrast, the spatial structure of the MBS wave functions is critical, particularly the (partial) separation of the two MBS wave functions. We note that in multi-band systems (topologically-trivial) low-energy ABSs that ``stick'' near zero energy can also be generated through the so-called inter-band coupling mechanism,\cite{Woods2019} in addition to the partial separation mechanism described here. In essence, band repulsion
resulting from strong inter-band coupling can pin the lowest energy state near zero energy over a significant range of control parameters. The resulting near-zero-energy ABSs are characterized by MBS components that are not spatially separated.\cite{Woods2019} In general, in multi-band systems the  inter-band coupling mechanism and the partial separation mechanism are expected to act in conjunction, with relative weights that depend on the details of the system and the values of the control parameters.

\subsubsection{Twisted Zeeman field}\label{SecIIIB3}

To strengthen our conclusion regarding the irrelevance  of the spin degree of freedom in the emergence of near-zero-energy ABS modes we now consider the rather artificial but conceptually clean case of a proximitized system with smooth confinement and a `twisted' Zeeman field. More specifically, we assume a position-dependent Zeeman field  given by
\begin{equation}
\Gamma(x) = \Gamma\times \tanh\left(\frac{L-2x}{2\delta_\Gamma}\right),  \label{Gmx}
\end{equation}
where $\delta_\Gamma=50~$nm defines a narrow transition region between the left half of the wire (characterized by a Zeman field $+\Gamma$) and its right half (which experiences a Zeeman field $-\Gamma$).
The system has soft confinement at both ends given by Gaussian barriers with $V_{max}=8~$meV and $\delta_{\rm v}=0.3~\mu$m.
For comparison, we also consider a system with uniform Zeeman field and soft confinement either at one end, or at both ends.

\begin{figure}[t]
\begin{center}
\includegraphics[width=0.49\textwidth]{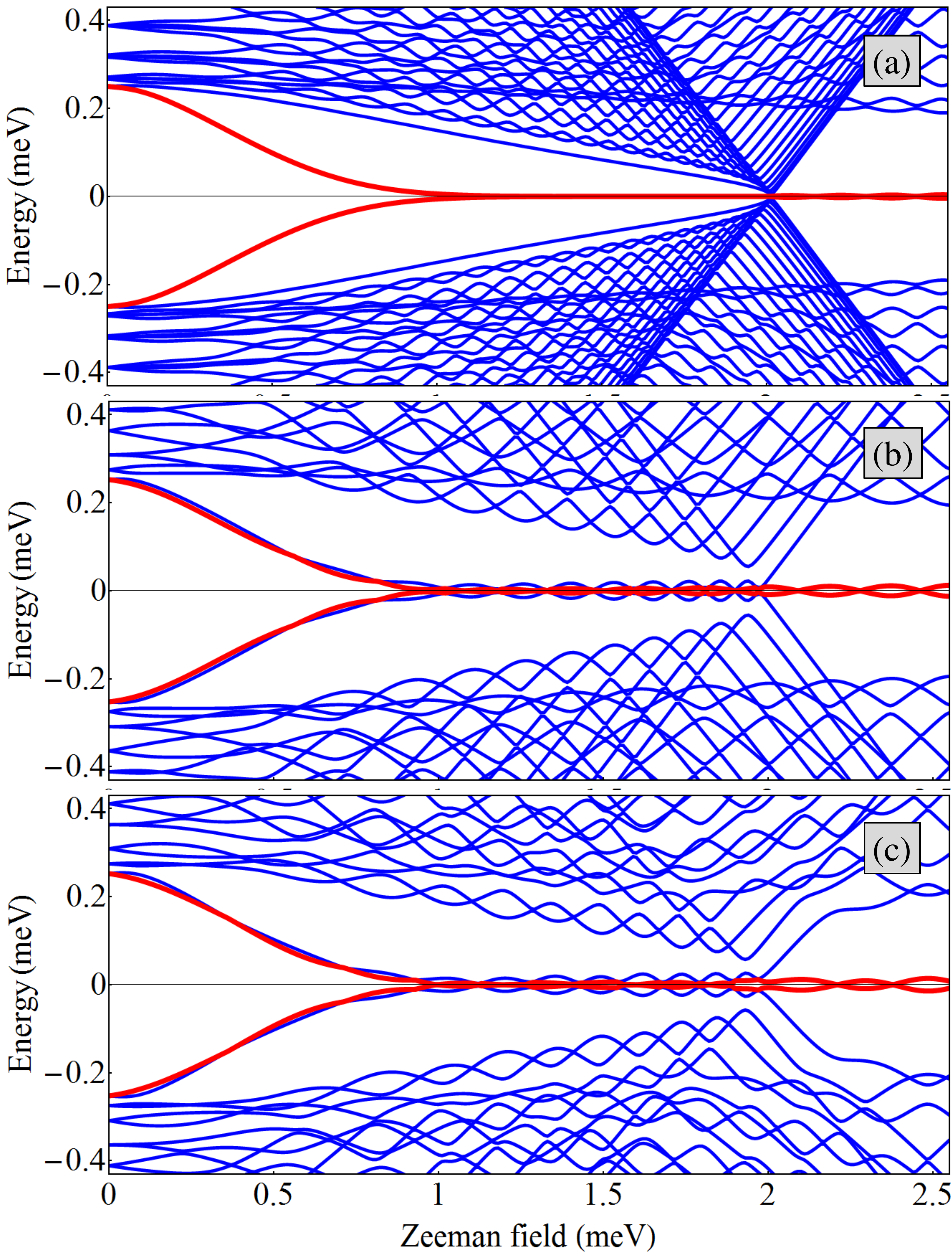}
\end{center}
\caption{Dependence of the low-energy spectrum on the strength of the applied Zeeman field for: (a) system with uniform Zeeman field and soft confinement at one end of the wire, (b)  system with uniform Zeeman field and soft confinement at both ends, and c)  system with twisted Zeeman field and soft confinement at both ends. The chemical potential is $\mu=2~$meV and the confinement potential is characterized by $V_{max}=8~$meV and $\delta_{\rm v}=0.3~\mu$m.}
\label{FG20}
\end{figure}

The dependence of the low-energy spectrum on the strength of the applied Zeeman field for these three scenarios is shown in Fig. \ref{FG20}.
In panel (a) we show the spectrum of a system with uniform Zeeman field and soft confinement at the left end of the wire (the system having  hard-wall confinement at the right end). This is an example of a system that supports an i-ABS at the right end of the wire (finite energy in-gap mode for $\Gamma <2~$meV) and a potential-induced ps-ABS at the left end (red line), similar to the situation discussed in the context of Figs. \ref{FG1}-\ref{FG4}. Note that, after collapsing at zero energy, the ps-ABS mode does not show visible energy splitting oscillations.
In panel (b) we show the spectrum of a system with uniform Zeeman field and soft confinement at both ends of the wire. The effective length of the wire is shorter than in panel (a). Note that for $\Gamma <\Gamma_c$ there are two near-zero-energy modes, while the characteristic i-ABS line (visible in panel (a) close to the bulk gap edge) is absent. The second-lowest mode (blue line) exhibits significant energy-splitting oscillations.
Finally, in panel (c)  we show the spectrum of a system with a twisted Zeeman field given by Eq. (\ref{Gmx}) and soft confinement at both ends of the wire. Note that the spectrum is qualitatively indistinguishable from that shown in panel (b). In particular, there are two near-zero-energy modes, with the second-lowest mode (blue line) exhibiting significant energy-splitting oscillations.

\begin{figure}[t]
\begin{center}
\includegraphics[width=0.49\textwidth]{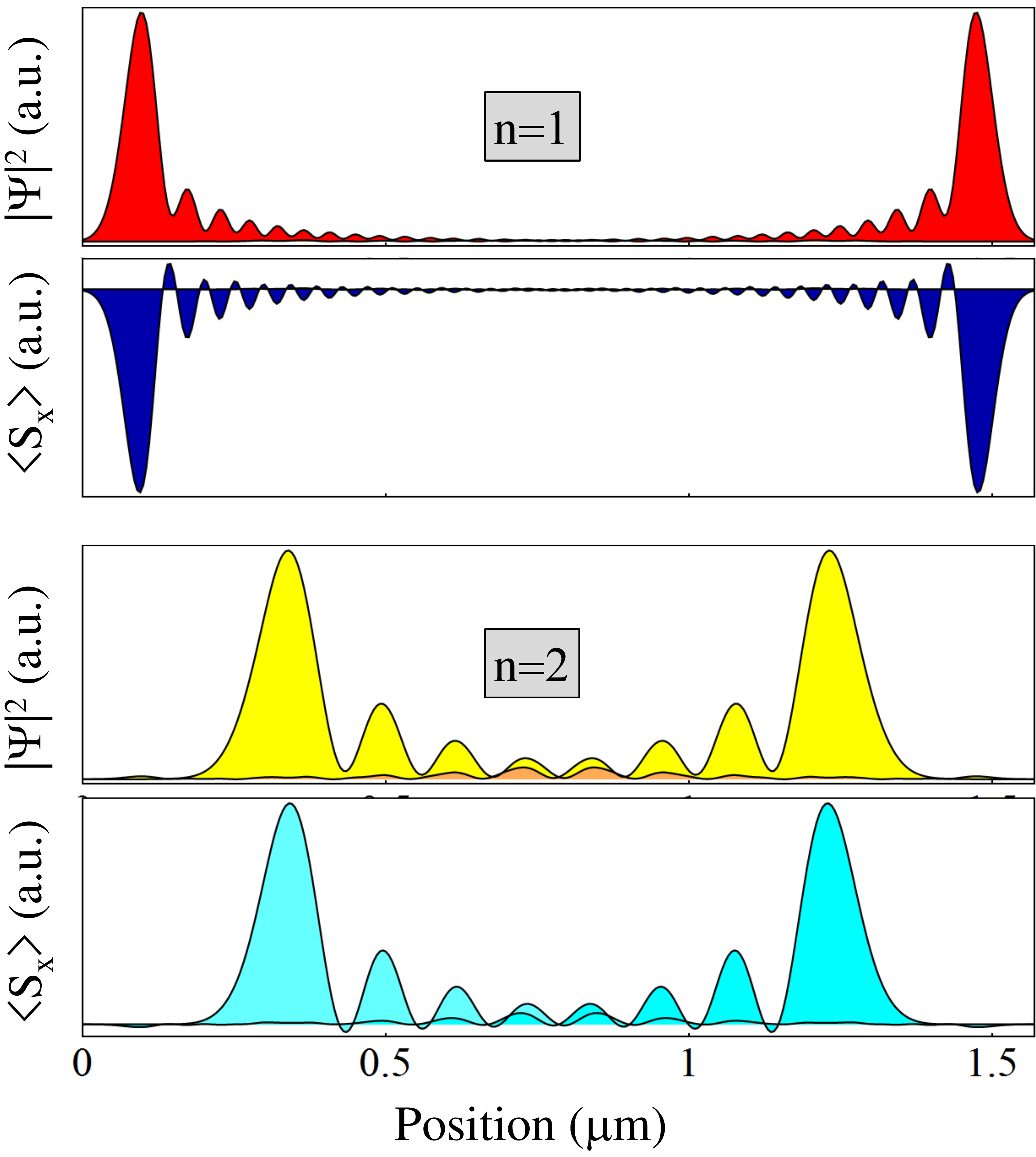}
\end{center}
\caption{{\em Top panels}: Majorana wave functions $\psi_A$ and $\psi_B$ [given by Eqs. (\ref{chiA} and\ref{chiB})] associated with the  lowest-energy mode ($n=1$) in Fig. \ref{FG20}(b) for a uniform Zeeman field $\Gamma=1.2~$meV and the corresponding spin density. We have used the same shading (red) for both Majoranas to suggest their association  with the same (spin-down) spin-split sub-band.{\em Bottom panels}: Majorana wave functions associated with the second  lowest energy mode ($n=2$) in Fig. \ref{FG20}(b) for $\Gamma=1.2~$meV and the corresponding spin density.}
\label{FG21}
\end{figure}

For a system with a uniform Zeeman field, the smooth confinement potential at the ends of the wire separates spatially the component MBSs of the corresponding in-gap ABSs, with the spin-down Majoranas being closer to the ends and the spin-up Majoranas being pushed toward the center of the wire. The well-separated spin-down MBSs (red shading in Fig. \ref{FG21}) form the lowest-energy fermionic mode ($n=1$), while the partially-overlapping spin-up MBSs (yellow) combine into the second-lowest-energy mode ($n=2$).
\begin{figure}[t]
\begin{center}
\includegraphics[width=0.49\textwidth]{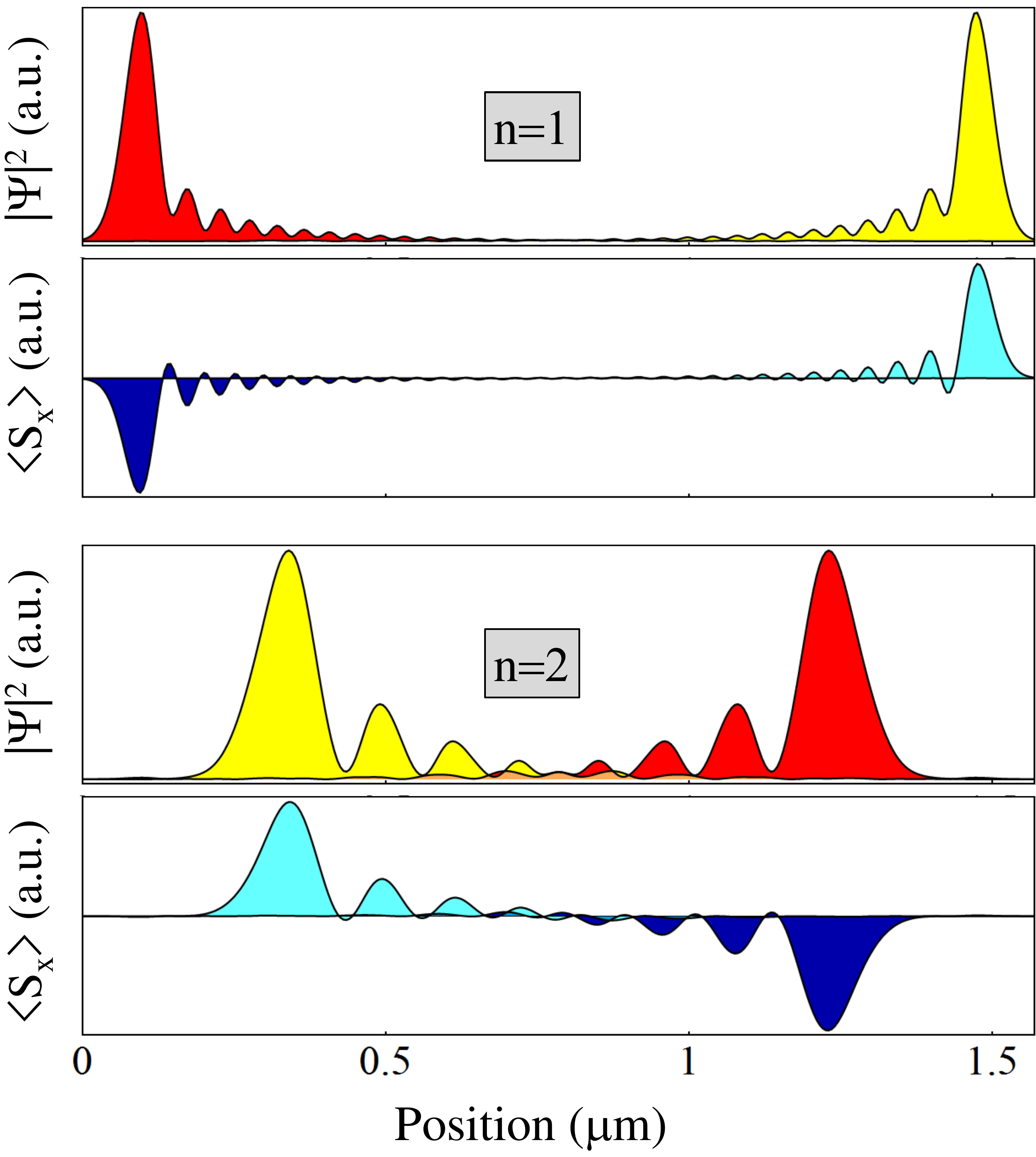}
\end{center}
\caption{{\em Top panels}: Majorana wave functions associated with the  lowest-energy mode ($n=1$) in Fig. \ref{FG20}(c) for a twisted Zeeman field with $\Gamma=1.2~$meV and the corresponding spin density. We have used different shading (red and yellow) for the two Majoranas to suggest their association  with different (spin-down and spin-up, respectively) spin-split sub-band.{\em Bottom panels}: Majorana wave functions associated with the second  lowest energy mode ($n=2$) in Fig. \ref{FG20}(c) for $\Gamma=1.2~$meV and the corresponding spin density.}
\label{FG22}
\end{figure}
Note that there is also a significant overlap between the spin-down MBSs (Fig. \ref{FG21}, top panels) and the corresponding spin-up Majoranas (Fig. \ref{FG21}, bottom panels), but this does not result in energy splitting, as shown by the spectrum in Fig. \ref{FG20}(a) (or, more clearly, by increasing the length of the wire, which does not affect the overlap between corresponding ``red'' and ``yellow'' Majoranas, yet generates a spectrum without visible energy-splitting oscillations). By contrast, the overlap between the exponential tails of the spin-up (yellow) Majoranas  (which point toward each other) results in the energy splitting oscillations that characterize the mode $n=2$ in Fig. \ref{FG20}(b). This mechanism is similar to that responsible for the Majorana energy splitting oscillations in clean, finite wires above the finite-size remnant of the TQPT.  We emphasize that, in contrast to this behavior, partially-overlapping MBSs characterized by exponential tails pointing in the same direction have a suppressed energy-splitting and no well-defined oscillations. This is the case for both MBSs with opposite spin character [e.g., generated according to scenario ($\alpha$)] and MBSs having the same spin character (see Sec. \ref{SecIIIB2}).

We now ask the key question of whether the energy splitting oscillations characterizing mode $n=2$ are the result of the spatial profile of the overlapping MBSs (specifically, the fact that the exponential tails are pointing toward each other, rather than pointing in the same direction or away from each other), or the result of the two MBSs being associated with the same spin-split sub-band (i.e., having the same sign of the spin density). To address this question, we analyze the spatial profiles of the Majorana wave functions associated with the lowest two (trivial) modes in Fig. \ref{FG20}(c), i.e. the modes characterizing a system with twisted Zeeman field and $\Gamma < \Gamma_c$. The results are presented in Fig. \ref{FG22}. Note that the spatial profiles of the MBSs are virtually identical to those in Fig. \ref{FG21}, while the spin character of the MBSs localized in the right half of the wire is reversed. We conclude that the presence/absence of energy-splitting oscillations is determined by the real-space properties of the MBS wave functions and not by their spin structure.

\begin{figure}[t]
\begin{center}
\includegraphics[width=0.49\textwidth]{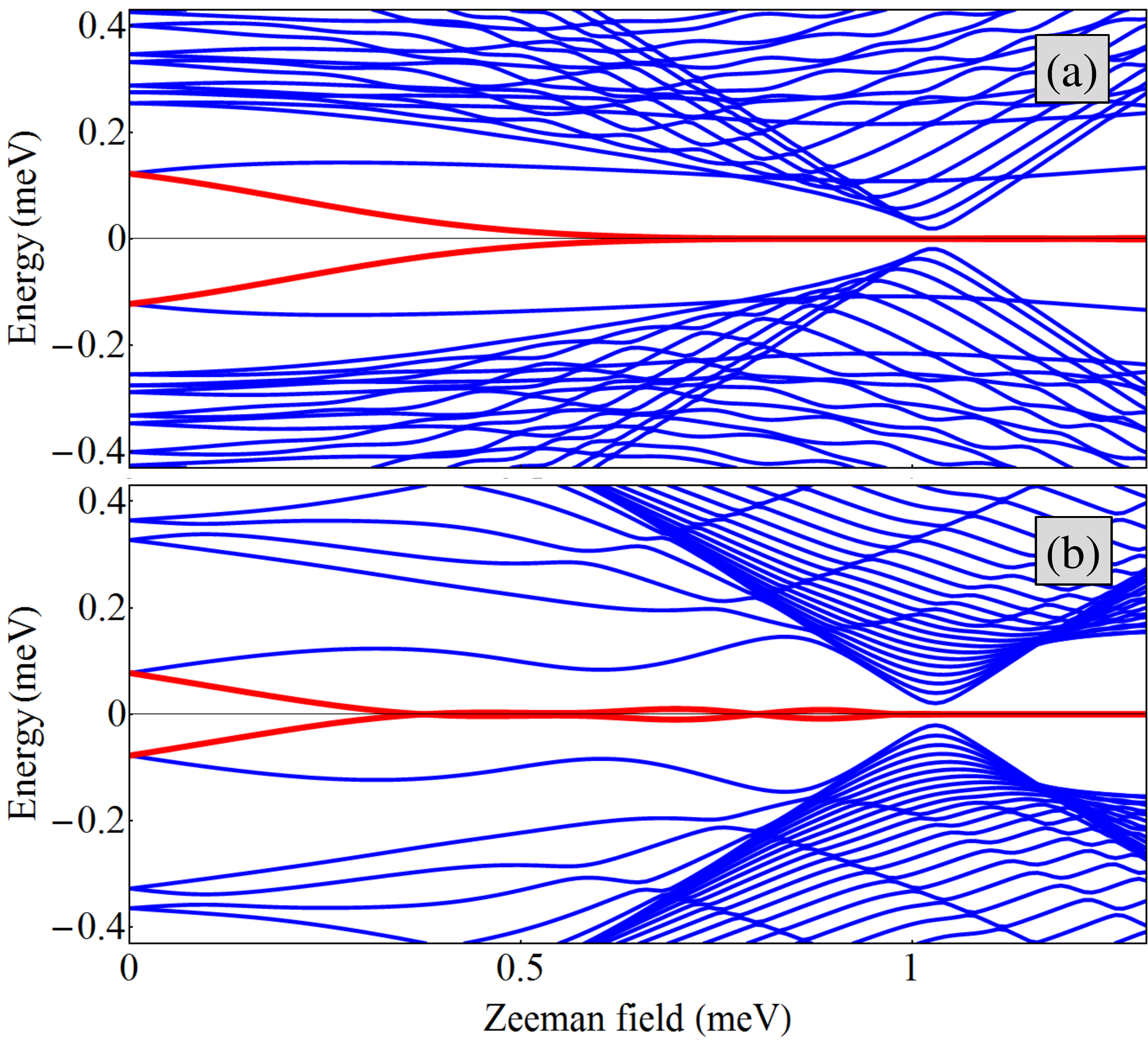}
\end{center}
\caption{(a) Dependence of the low-energy spectrum on the applied Zeeman field for a system with effective potential given by Eq. (\ref{VV2}) with $V_0=2.3~$meV and chemical potential $\mu=1~$meV.  (b) Low-energy spectrum for a system with a quantum well defined by Eq. (\ref{VV2}) with  $V_0=-2.3~\mu$m. The chemical potential  is $\mu=-1~\mu$m.}
\label{FG23}
\end{figure}

In conclusion, there is direct correlation between increasing the separation of the component MBSs of a ps-ABS and the collapse of its energy toward zero. This property is clearly illustrated by the results shown in Fig. \ref{FG11} and Figs. \ref{FG17} and \ref{FG19}.  In addition, our results show that the spin structure of the corresponding MBSs play no role in this collapse, hence in the emergence of robust near-zero energy ps-ABSs (see, for example, the results in Fig. \ref{FG18} and Figs. \ref{FG21}-\ref{FG22}).  We note that fully overlapping MBSs have, in general, finite energies comparable to the induced gap, as illustrated  by the intrinsic ABS modes shown in Fig. \ref{FG12}. This statement applies to hybrid systems with well-separated bands, which are the object of this study, but not to coupled multi-band systems, where trivial ABSs can emerge due to inter-band mixing.\cite{Woods2019} 

\section{Charge tunneling signatures}\label{SecIV}
An experimentally-relevant aspect that we want to address is the relationship between the properties of the component MBSs and the signatures of a low-energy ABS in a local charge tunneling measurement. We consider the case of a wire with a position-dependent potential and a short uncovered region at the left end (see Fig. \ref{FG9}). The  total length of the wire is $L=2~\mu$m, while the length of the uncovered region is $L_\Delta= 0.3~\mu$m. The potential at the (left) end of the wire has the form
\begin{equation}
V(x) = \frac{V_0}{2}\left[\tanh\left(\frac{x_0-x}{\delta_{\rm v}}\right) +1\right], \label{VV2}
\end{equation}
with $x_0=0.25~\mu$m and $\delta_{\rm v}=0.2~\mu$m. We also assume a narrow barrier between the left end of the SM wire and the normal lead, which is modeled as a weak link. This way the transparency of the barrier can be controlled (in part) independently of the position-dependent potential.
We calculate the differential conductance using the Blonders-Tinkham-Klawijk (BTK) formalism\cite{Blonder1982} for a wire [described by Hamiltonian (\ref{H}) with a potential given by Eq. (\ref{VV2})] coupled to a normal lead (described by a tight-binding model with nearest-neighbor hopping).\cite{Lin2012,Stenger2017} 

We investigate two different cases: (i) a system with smooth confining potential corresponding to $V_0=2.3~$meV in Eq. (\ref{VV2}) and (ii) a system with an asymmetric potential well described by Eq. (\ref{VV2})  with $V_0=-2.3~$meV.
\begin{figure}[t]
\begin{center}
\includegraphics[width=0.49\textwidth]{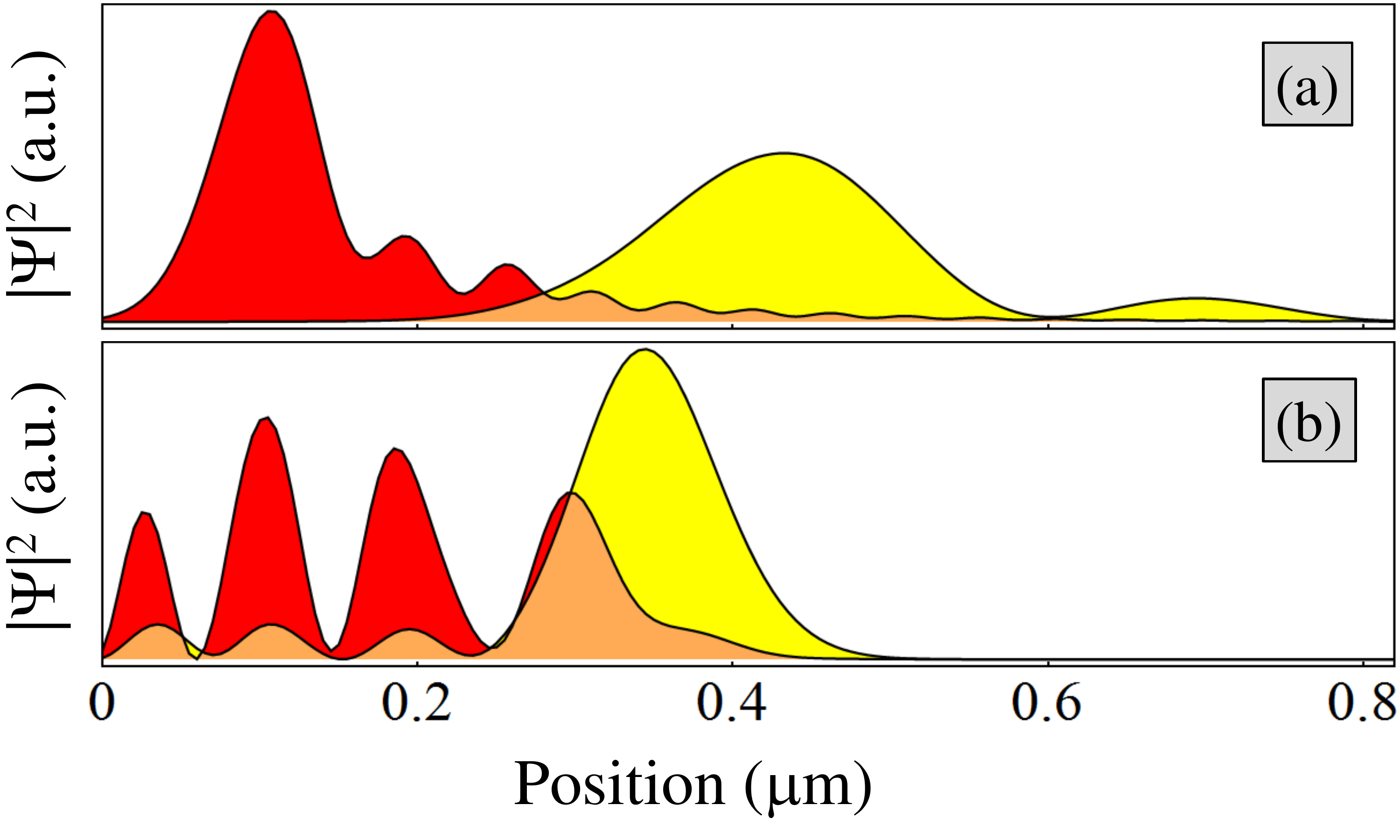}
\end{center}
\caption{(a) Majorana wave functions associated with the near-zero-energy ABS mode from Fig. \ref{FG23}(a) corresponding to $\Gamma=0.75~$meV. The two-component MBSs have (preponderantly) opposite spin character. (b) Majorana wave functions associated with the near-zero-energy ABS mode from Fig. \ref{FG23}(b) corresponding to $\Gamma=0.5~$meV. The ``yellow'' MBS is associated with the spin-up sub-band, while the ``red'' Majorana has mixed spin character.}
\label{FG24}
\end{figure}
The dependence of the corresponding low-energy spectra on the applied Zeeman field is shown in Fig. \ref{FG23}.
The ps-ABS mode in panel (a) is generated by a ``standard'' smooth confinement mechanism. By contrast, the low-energy mode in panel (b) is an example of ABS generated by a ``mixed''  mechanism in an asymmetric quantum well that is about half-filled/empty.
Note that the energy of both ABS modes at $\Gamma=0$  is significantly lower than the induced gap ($\Delta =0.25~$meV). This is a clear signature of the ABS being (partially) localized outside the proximitized segment of the wire.

As shown in Fig. \ref{FG24} panel (a), in the system with soft confinement the low-energy ABS consists of two well separated MBSs that will couple very differently to a local probe at the left end of the wire. Basically, one will only couple to the ``red'' Majorana, while the ``yellow'' Majorana will remain virtually invisible. By contrast, in the system with a potential well [panel (b)] the component Majorana bound state localized further away from the left end (i.e., the ``yellow'' MBS) has a tail that points toward the end of the wire, which ensures a reasonably good coupling to the local probe.

\begin{figure}[t]
\begin{center}
\includegraphics[width=0.49\textwidth]{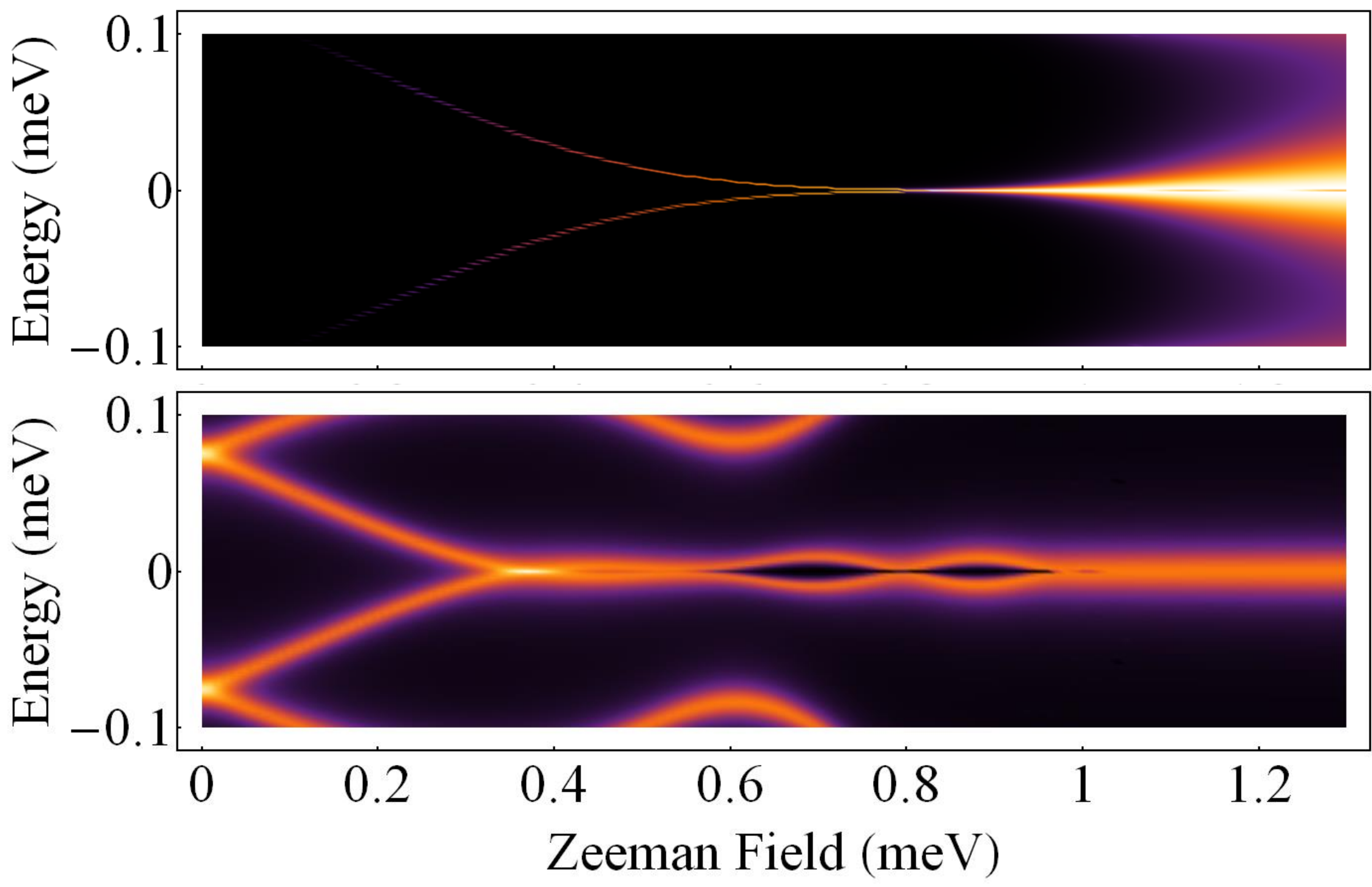}
\end{center}
\caption{(a) Zero-temperature differential conductance as function of energy and Zeeman field for a system with smooth barrier. The system parameters are the same as in Fig. \ref{FG23}(a). The zero-bias conductance peak is quantized at $2e^2/h$, but its width increases strongly with the Zeeman field. There is no signature associated with the TQPT at $\Gamma_c\approx 1~$meV. (b) Zero-temperature differential conductance as function of energy and Zeeman field for a system with a potential well. The system parameters are the same as in Fig. \ref{FG23}(b). The width of the low-energy peaks is weakly dependent on the Zeeman field, the height is not quantized for $\Gamma<\Gamma_c$ (also see Fig. \ref{FG26}).}
\label{FG25}
\end{figure}

\begin{figure}[t]
\begin{center}
\includegraphics[width=0.49\textwidth]{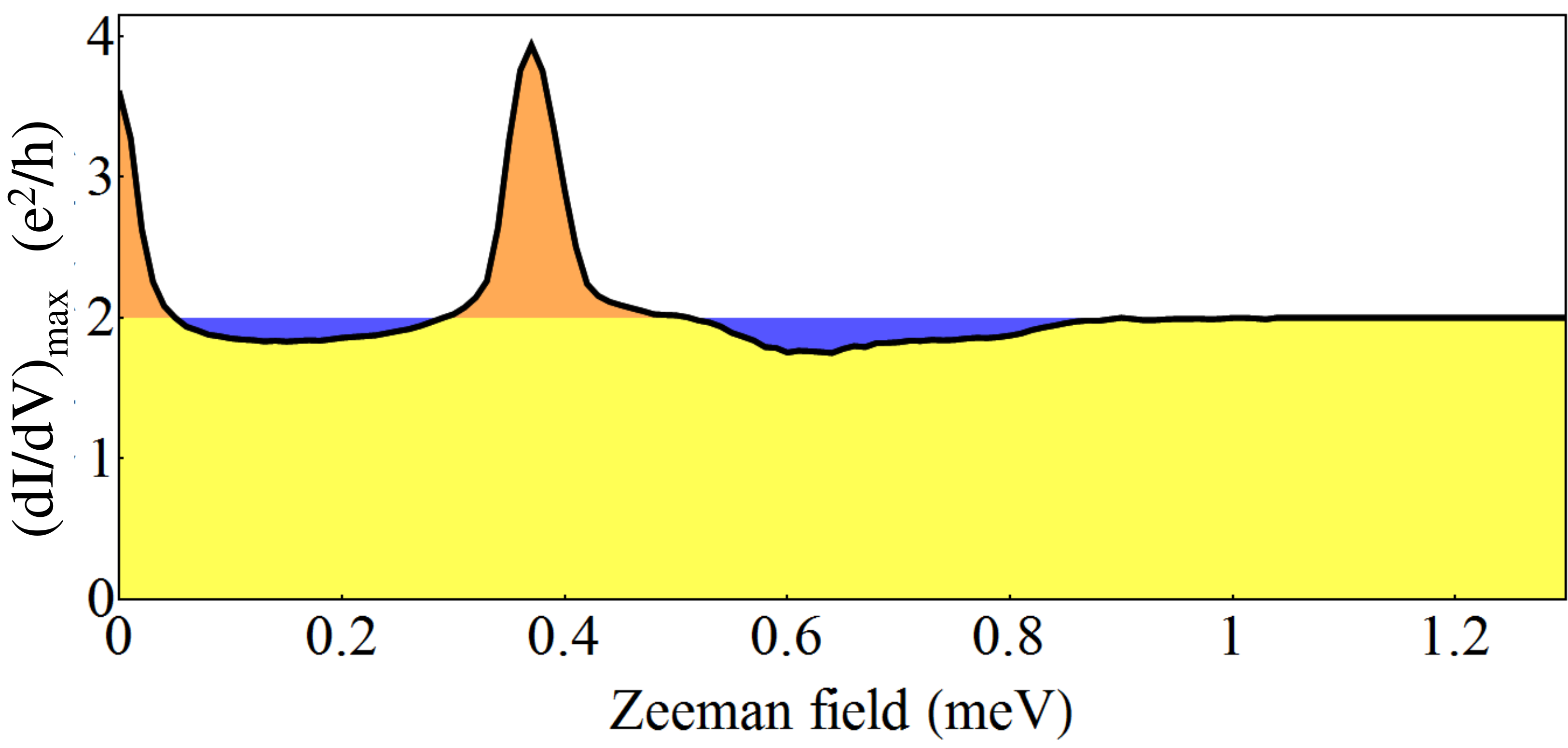}
\end{center}
\caption{Height of the lowest-energy conductance peaks showed  in Fig. \ref{FG25}(b) as function of the Zeeman field. Note that the zero-temperature  differential conductance peak is not quantized below the critical field  $\Gamma_c\approx 1~$meV.}
\label{FG26}
\end{figure}

The dependence of the  differential conductance on the applied Zeeman field and the bias voltage (energy) is shown in Fig. \ref{FG25}.
The low-energy conductance peak for the system with smooth confinement [Fig. 25 panel (a)] is quantized to $2e^2/h$ even before the ps-ABS mode collapses to zero energy. This is consistent with the profiles of the component MBSs shown in Fig. \ref{FG24}(a), which suggest that only the `red' Majorana couples (measurably) to the local probe at the end of the wire.
The width of the conductance peak increases strongly with the applied Zeeman field. Note that  for $\Gamma < 0.7~$meV the width of the conductance peak is comparable to finite grid used in the plot (and the peak becomes barely visible despite its height being $2e^2/h$ at zero temperature).
Of course, any finite temperature would result in the complete disappearance of the conductance peak at low fields. 
  The strong dependence on the Zeeman field is a characteristic of ps-ABSs induced by the barrier potential itself, since the height of the ``effective barrier'' experienced by the strongly coupled (``red'')  MBS decreases   linearly with $\Gamma$.
By contrast, the low-energy conductance peak for the system with a potential well [Fig. 25 panel (b)] is {\em not} quantized for $\Gamma < \Gamma_c$ (see also Fig. \ref{FG26}). This is a consequence of both component MBSs having  measurable coupling to the local probe.  Note that having a relatively robust low-energy mode that sticks to zero energy over a large range of Zeeman fields does not guarantee the quantization of the corresponding conductance peak to a height $2e^2/h$.
On the other hand, the width of the conductance peak is weakly dependent on the Zeeman field because the transparency of the barrier is controlled by the weak link, rather than the position-dependent potential.
Note also that the visible minimum gap at $\Gamma\approx 0.6~$ meV (see Fig. \ref{FG25}, lower panel) has nothing to do with the bulk gap ``closing and reopening'' at the TQPT.  In fact, this feature is associated with a finite-energy ABS [see Fig. \ref{FG23}(b)] localized near the end of the wire.

Finally, we note that, unlike propagating modes the coupling of bound states to the normal leads is not simply determined by the potential barrier (e.g., its height and width), but depends critically on the ``location'' of the bound state, i.e. the spatial properties of its wave function. For example, a bound state localized far from the tunnel barrier will be weakly coupled to the lead (and practically ``invisible'' in a tunneling conductance measurement) regardless of the tunnel barrier height. This key property provides a simple explanation of the results shown in Figs. \ref{FG25} and \ref{FG26}. For example, the specific shape of the MBS wave functions shown in Fig. \ref{FG24}(b) ensures that both MBSs couple (significantly) to the lead and, consequently, the corresponding nearly-zero-energy conductance peak is not quantized in the topologically trivial regime (see Fig. \ref{FG26}). Furthermore, the example shown in Fig. \ref{FG14} corresponds to a spin-up (yellow) MBS that couples more strongly to a (left) lead than the spin-down (red) MBS, despite the fact that the spin-up effective potential barrier is higher than  the spin-down barrier. This shows clearly that the coupling to the leads is ultimately controlled by the spatial properties of the component MBSs (including their spatial separation) and not by their spin structure.

\section{Stability of ps-ABS modes in the presence of disorder}\label{SecV}

The  near-zero energy ABSs emerging in the topologically trivial regime through the partial separation mechanism discussed here are not topologically protected. Partial separation of the component MBSs implies partial overlap, hence finite sensitivity to local perturbations. The natural questions are how robust are these near-zero energy states in the presence of local perturbations, e.g., various types of disorder, and what is the best way to describe the effect of such perturbations?  However, before addressing these questions, we would like to clarify a few aspects regarding the terminology used to describe the low-energy modes in semiconductor-superconductor hybrid systems, in particular the relation between ps-ABSs (or quasi-Majoranas) and Majorana zero modes (MZMs).  We propose the following operational distinction. Consider the characteristic length scale $L^*(\Gamma)$ defined in Sec. \ref{SecII} corresponding to some typical value of the Zeeman field associated with the presence of (near) zero energy modes. According to our analysis, $L^*(\Gamma)$ determines the separation of the Majorana modes. Let us first first assume a long wire with $L \gg \xi_M$,  where $L$ is the length of the wire and $\xi_M$ is the characteristic Majorana length scale. If $L^* \sim \xi_M$, the low-energy state is a ps-ABS, while  $L^*\gg \xi_M$ corresponds to well-separated, ``genuine'' MZMs. For short wires ($L \gtrsim \xi_M$),  $L^* \lesssim L/2$ corresponds to a ps-ABS localized near one end of the wire, while $L^*\sim L$ implies the presence of ``precursor'' MZMs localized at the opposite ends of the system. A few remarks are warranted. First, we note that the rationale for the distinction proposed above in the case of short wires  is based on the following observation: in uniform-enough systems the near-zero energy modes  are MBSs localized near the ends of the wire (which can be viewed as ``precursor'' MZMs), while inhomogeneities can generate a pair of MBSs localized near one end (hence, a ps-ABS). Second, in long wires the MZMs are not necessarily localized at the ends of the system, but at the ends of a segment of length $L^*\leq L$ that can be viewed as topologically nontrivial. Note that in the case of potential shores, which support type-($\alpha$) MBSs, one can consider a ``thermodynamic limit'' corresponding to $L\rightarrow \infty$ and $\Delta V={\rm const.}$, where $\Delta V$ is the difference between the maximum and the minimum of the effective potential throughout the wire. This limit corresponds to a vanishing average slope of the effective potential, i.e., a nearly-uniform infinite system. In this limit, the system supports a pair of type-($\alpha$)  MBSs associated with {\em different} spin sub-bands that are separated by a distance  $\sim L\rightarrow\infty$.  Finally, we emphasize that a ps-ABS ($L^* \sim \xi_M$) can be continuously connected to a ``genuine'' MZM ($L^*\gg \xi_M$) by continuously varying the system parameters. In practice, the relevant questions are whether or not this ``transition'' can be performed in a controllable manner and how protected are the (near-) zero-energy modes?

\begin{figure}[t]
\begin{center}
\includegraphics[width=0.49\textwidth]{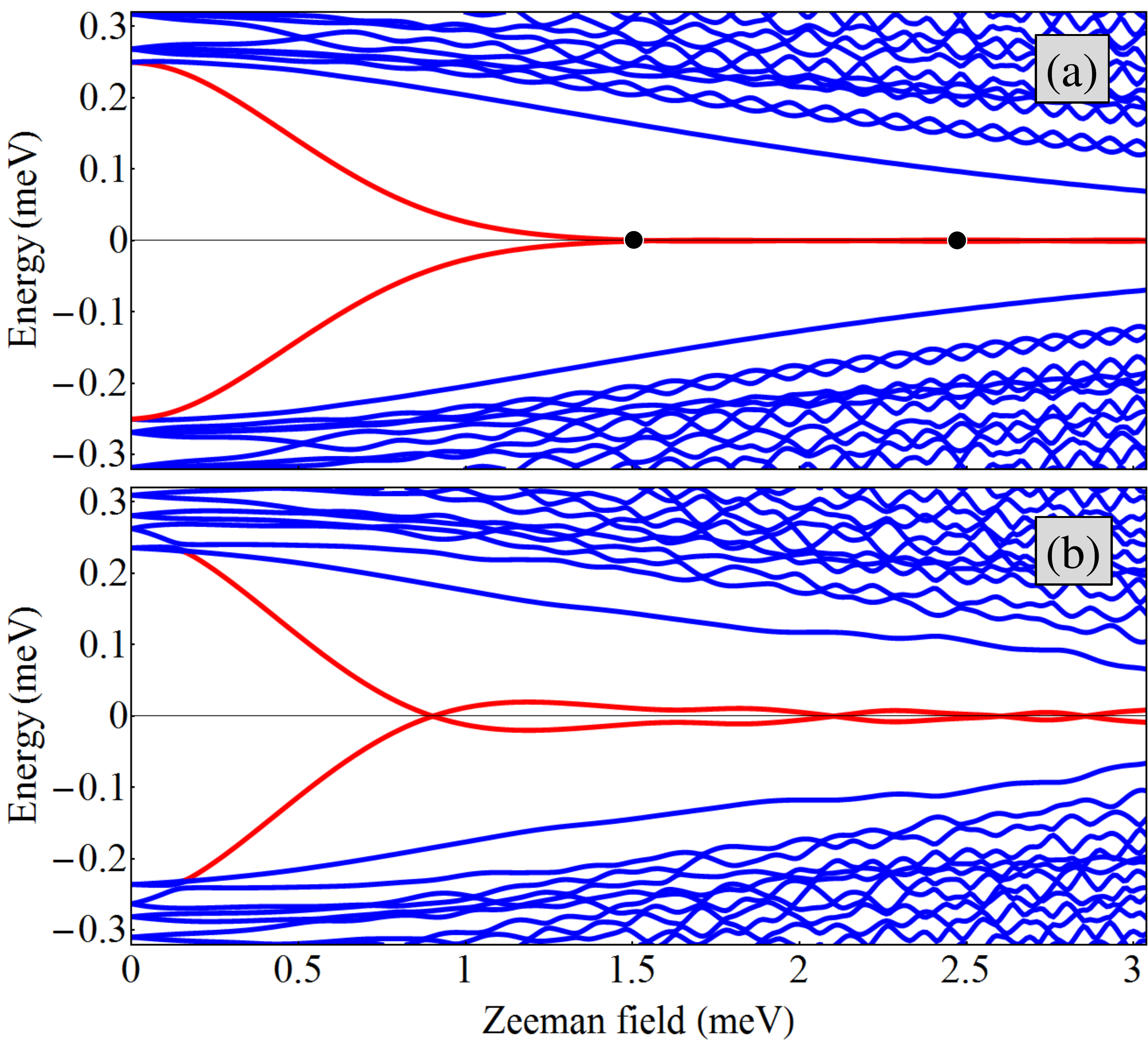}
\end{center}
\caption{(a) Dependence of the low-energy spectrum on the applied Zeeman field for a clean system with chemical potential $\mu=5~$meV and effective potential given by Eq. (\ref{Vx1}) with $\kappa=20~$meV$/\mu$m and $L=3~\mu$m. The lowest-energy mode (red lines) corresponds to a ps-ABS; the wave functions of the states marked by black dots are shown in Fig. \ref{FG28}. (b) Same as in panel (a) for a disordered system with potential and pairing disorder described by Eq. (\ref{Hdis}) with $\delta V = 0.5~$meV and $\delta \Delta = -0.25~$meV. The corresponding disorder profiles are shown in Fig. \ref{FG29}, panels (a) and (b), respectively. Note that in the presence of disorder the low-energy ps-ABS mode acquires an energy splitting on the order of $20~\mu$eV.}
\label{FG27}
\end{figure}

\begin{figure}[t]
\begin{center}
\includegraphics[width=0.49\textwidth]{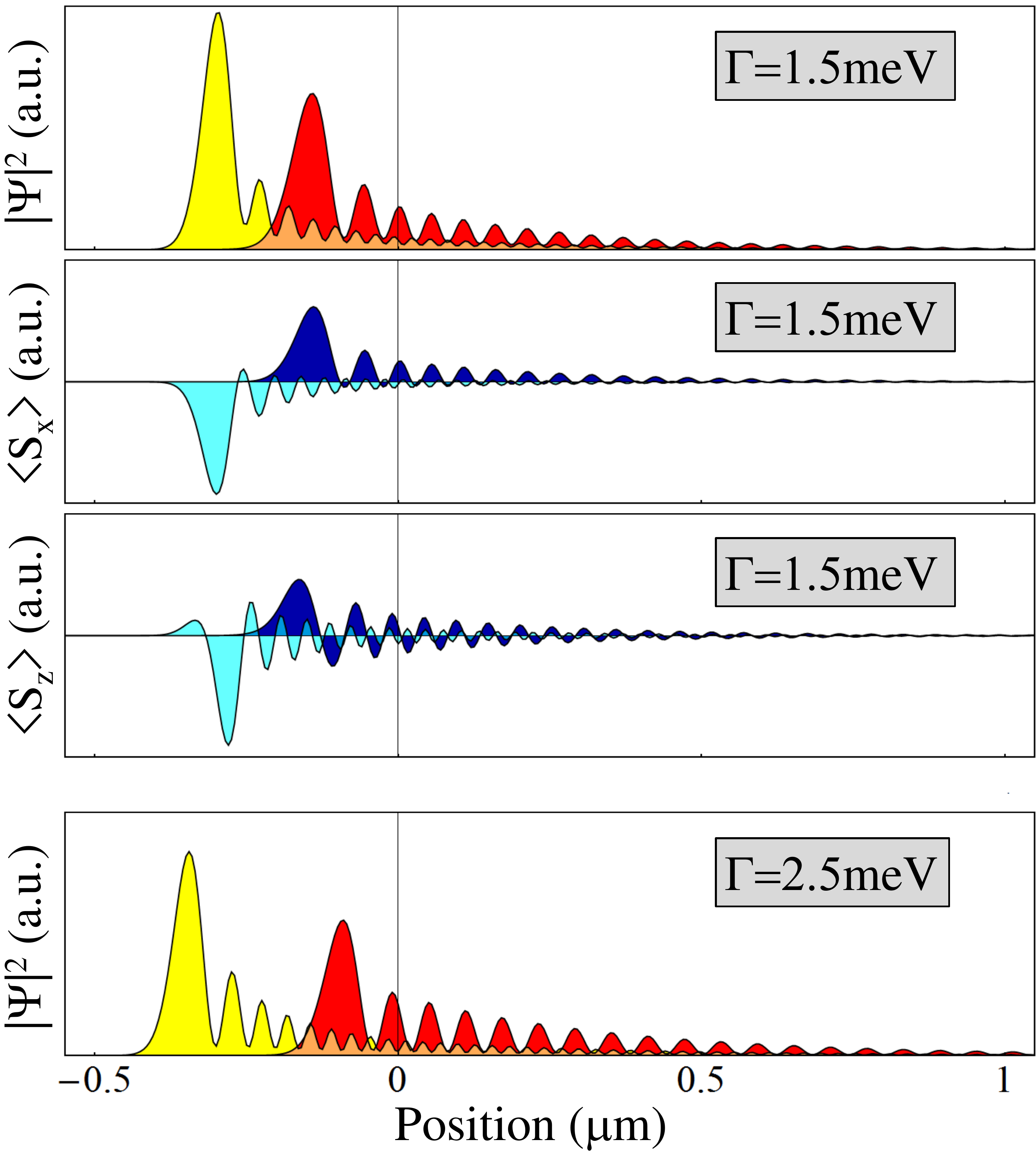}
\end{center}
\caption{Top and bottom panels: Majorana wave functions corresponding to the ps-ABSs marked by black dots in Fig. \ref{FG27}(a). Note that the separation of the MBS components increases with the Zeeman field. Middle panels: Spin densities of the Majorana components along the x and z directions. We note that $\langle S_y\rangle = 0$.}
\label{FG28}
\end{figure}

Consider a system with an effective potential described by Eq. (\ref{Vx1}) with a slope $\kappa= 20~$meV/$\mu$m and length $L=3~\mu$m. We fix the chemical potential at $\mu=5~$meV, which corresponds to a critical Zeeman field $\Gamma_c \approx 5~$meV. In the absence of disorder, the system supports a low-energy ps-ABS mode at Zeeman fields much lower than the critical value, as shown in the upper panel of Fig. \ref{FG27}. The wave functions of the Majorana components of the ABS for Zeman fields $\Gamma=1.5~$meV and $\Gamma=2.5~$meV [black dots in Fig. \ref{FG27}(a)] are shown in Fig. \ref{FG28}, as well as the spin polarizations along the x- and z-axes for the ps-ABS at $\Gamma=1.5~$meV. First, note the characteristic spatial separation of the component MBSs, which increases with increasing Zeeman field. Second, let us note that the corresponding spin polarizations are nonzero along both x and z directions, while $\langle S_y\rangle = 0$. While the spin densities $\langle S_x\rangle$ corresponding to the two MBSs have opposite signs almost everywhere -- reflecting the association of the two MBSs with different spin sub-bands -- this correlation between spin density and  spin sub-bands is less manifest in the case of $\langle S_z\rangle$. Consequently, any statement regarding ($\alpha$)-type MBSs components having almost  ``opposite spin'' should be understood as referring primarily to $\langle S_x\rangle$ or, more precisely,  the association of the two MBSs with different spin sub-bands. 

\begin{figure}[t]
\begin{center}
\includegraphics[width=0.49\textwidth]{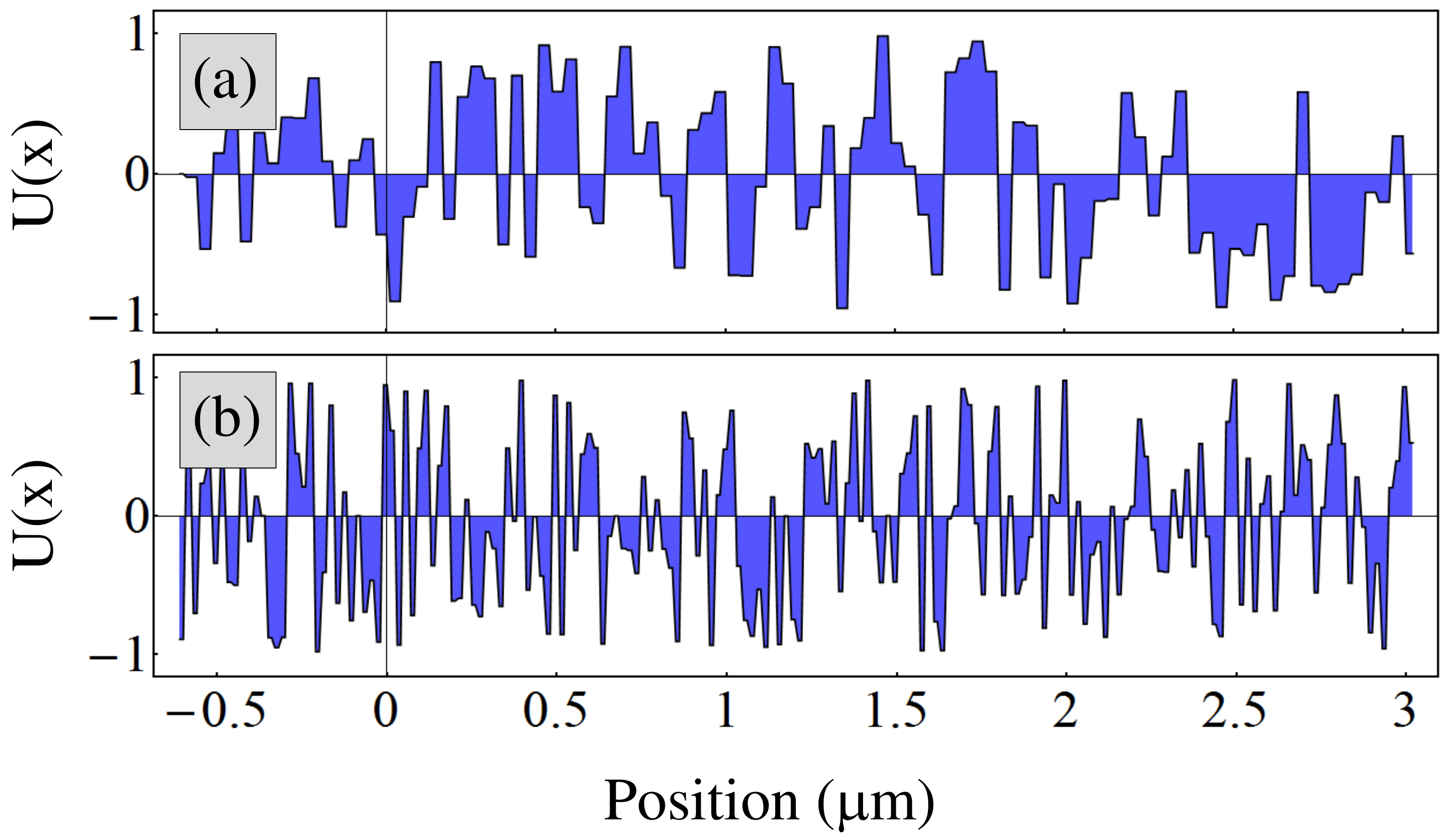}
\end{center}
\caption{Disorder profiles corresponding to (a) $\delta_U=40~$nm and (b) $\delta_U=20~$nm. The function $U(x)$  has constant but random values $-1 \leq U(x) \leq 1$ within each segment $\delta_U$ of a $3.6~\mu$m long wire.}
\label{FG29}
\end{figure}

To investigate the effect of disorder, we supplement the BdG Hamiltonian with a disorder-induced term of the form $H_{\rm dis} = \sum_i \Psi_i^\dagger {\cal H}_{\rm dis}(i) \Psi_i$, where $\Psi^\dagger_i = (c_{i\uparrow}^\dagger, c_{i\downarrow}^\dagger, c_{i\uparrow}, c_{i\downarrow})$ is a Nambu operator acting on site $i$ and the (first quantized) Hamiltonian ${\cal H}_{\rm dis}$ is given by
\begin{eqnarray}
{\cal H}_{\rm dis}(x) &=& \delta V ~\!U_V\!(x)~\! \tau_z + \delta \Gamma_x ~\!U_{\Gamma_x}\!(x) ~\! \sigma_x\tau_z  + \delta \Gamma_y ~\!U_{\Gamma_y}\!(x) ~\! \sigma_y\nonumber \\
&+& \delta \Gamma_z~\! U_{\Gamma_z}\!(x)~\! \sigma_z\tau_z +\delta\Delta ~\! U_\Delta\!(x) ~\! \sigma_y\tau_y.                                          \label{Hdis}
\end{eqnarray}
Here, $\sigma_\alpha$ and $\tau_\alpha$ (with $\alpha = x, y, z$) are Pauli matrices associated with the spin and particle-hole degrees of freedom, respectively,  $U_\Lambda(x)$ are dimensionless disorder profiles corresponding to random effective potentials ($\Lambda = V$), Zeeman fields ($\Lambda = \Gamma_\alpha$), and pairing potentials  ($\Lambda = \Delta$), while  $\delta \Lambda$ are the corresponding disorder amplitudes. In this study, we focus on  a class of disorder profiles $U(x)$ having constant but random values $-1 \leq U(x) \leq 1$ within segments of length $\delta_U$. Two specific disorder realizations corresponding to $\delta_U=40~$nm and  $\delta_U=20~$nm are shown in Fig. \ref{FG29}, panels (a) and (b), respectively. 

Consider now the system characterized by the (clean) low-energy spectrum shown in Fig. \ref{FG27}(a), which supports a near-zero energy ps-ABS mode (see Fig. \ref{FG28}), in the presence of disorder. For concreteness, we assume that the system has potential disorder characterized by a disorder profile $U_V$ corresponding to panel (a) of Fig. \ref{FG29} and an amplitude $\delta_V=0.5~$meV and pairing disorder with a profile $U_\Delta$ corresponding to panel (b) of Fig. \ref{FG29} and an amplitude $\delta_\Delta=-0.25~$meV. Note that the disorder amplitudes are on the order of the induced SC pairing, more specifically, $\delta_V=2\Delta$ and  $\delta_\Delta=-\Delta$. The corresponding low-energy spectrum is shown in Fig. \ref{FG27}(b). The most significant effect of disorder is the splitting of the ps-ABS mode by an energy on the order of $20~\mu$eV. 

\begin{figure}[t]
\begin{center}
\includegraphics[width=0.49\textwidth]{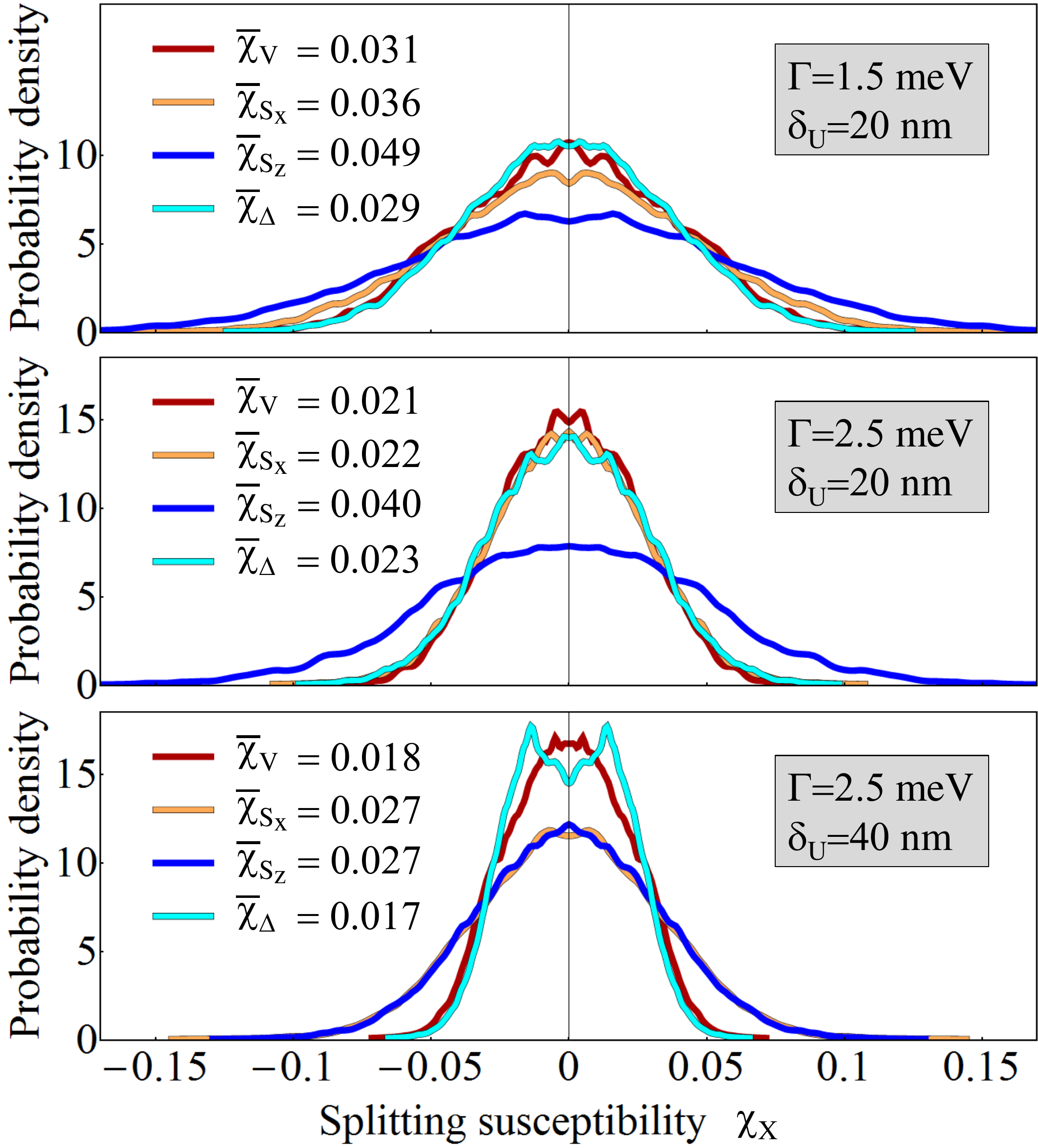}
\end{center}
\vspace{-3mm}
\caption{Probability distributions of the splitting susceptibilities defined by Eq. (\ref{chiL}) for the ps-ABS modes shown in Fig. \ref{FG28} in the presence of different types of disorder with profiles similar to those shown in Fig. \ref{FG29}. The results were obtained using 2000 randomly generated disorder profiles $U(x)$, as well as their ``flipped'' correspondents $-U(x)$. The ``average'' susceptibilities $\overline{\chi}_\Lambda$ are determined using Eq. (\ref{chiLL}).}
\vspace{-3mm}
\label{FG30}
\end{figure}

What does this simple example teach us about the robustness of the ps-ABS modes? On the one hand, if robustness is judged based on the signature of the ps-ABS mode in a charge tunneling experiment, we can conclude that the  low-energy mode is remarkably robust.  Indeed, considering the finite energy resolution of a differential conductance measurement (typically on the order of $10-20~\mu$eV), the system will still exhibit a robust zero-bias conductance peak that sticks to zero energy above a certain value of the Zeeman field. Moreover, since the Majorana wave functions of the ps-ABS are only weakly perturbed (as we checked explicitly), the two component MBS will couple asymmetrically to the external lead and the ZBCP will still be quantized  ($2e^2/h$) at low-enough temperatures (as in the clean case). Consequently, from this perspective, ps-ABSs are robust-enough and, basically, indistinguishable from bona fide MZMs. In the language of non-Hermitian topology, one can even view them as topologically nontrivial.\cite{Avila2018} On the other hand, from the perspective of topological quantum computation, the ps-ABS (as defined above) is not topologically protected and not suitable as a platform for a topological qubit. To quantitatively describe the effect of disorder,  let us project the total Hamiltonian onto the subspace define by the lowest-energy mode. In the Majorana basis $(\psi_A, \psi_B)$ defined by Eqs. (\ref{chiA} and \ref{chiB}) we have 
\begin{equation}
H_M = \left(
\begin{array}{cc}
0 &- i\epsilon \\
i\epsilon & 0
\end{array}\right),
\end{equation}
where $\epsilon=\epsilon_0+\delta\epsilon$ is the energy splitting in the presence of disorder. Here, $\epsilon_0 = i\langle\psi_A|H|\psi_B\rangle$, with $H$ given by Eq. (\ref{H}), is the energy splitting of the ps-ABS mode in a clean system and $\delta\epsilon = i\langle\psi_A|H_{\rm dis}|\psi_B\rangle$,  with $H_{\rm dis}$ given by Eq. (\ref{Hdis}), is the additional splitting due to the presence of disorder. For weak-enough disorder, the Majorana wave functions are weakly perturbed and we have 
\begin{equation}
\delta\epsilon \approx \sum_\Lambda \chi_\Lambda ~\delta\Lambda +\sum_{\Lambda,\Lambda^\prime}\chi_{\Lambda\Lambda^\prime}^{(2)} ~\delta\Lambda ~\!\delta\Lambda\prime +\dots,
\end{equation}
where the splitting susceptibilities can be calculated using the unperturbed Majorana wave functions, e.g., 
\begin{equation}
\chi_\Lambda =  i\langle\psi_A|H_{\rm dis}|\psi_B\rangle/\delta\Lambda\vert_{(\delta V,\dots,\delta\Delta)=(0,\dots,0)},  \label{chiL}
\end{equation}
with $(\psi_A, \psi_B)$ being the Majorana wave functions of the clean system. Note that $\chi_\Lambda$, with $\Lambda = V,\dots,\Delta$, depends on the control  parameters (e.g., the Zeeman field), as well as the disorder realization, $U_\Lambda$. To better characterize the effect of a certain {\em type of disorder} on the ps-ABS mode, it is useful to calculate the probability distribution of the corresponding splitting susceptibility. As an example, we consider disorder profiles  $U(x)$  having  constant but random values $-1 \leq U(x) \leq 1$ within segments of length $\delta_U$ (see Fig. \ref{FG29}). The corresponding  splitting susceptibilities for potential ($V$),  spin ($\Gamma_\alpha$), and pairing ($\Delta$) disorder are shown in Fig. \ref{FG30}. Note that $\chi_{\Gamma_y}=0$ and was not included. The probability distributions shown in Fig. \ref{FG30} were obtained using 2000 randomly generated disorder profiles $U(x)$, as well as their ``flipped'' correspondents $-U(x)$. A convenient parameter that describes a probability distribution is the ``average'' susceptibility
\begin{equation}
\overline{\chi}_\Lambda = \int |\chi_\Lambda|~{\cal P}_\Lambda (\chi_\Lambda)~ d\chi_\Lambda,  \label{chiLL}
\end{equation} 
where $ {\cal P}_\Lambda$ is the corresponding probability density. Examining the results shown in Fig. \ref{FG30}, we notice that the the ``average''  splitting susceptibilities at $\Gamma=2.5~$meV are significantly smaller than the corresponding susceptibilities at $\Gamma=1.5~$meV. This is consistent with the increase of the spatial separation of the component MBSs  (and the corresponding reduction of the Majorana wave functions overlap) shown in Fig. \ref{FG28}. We also notice the dependence  on the characteristic length scale of the disorder profile. For example, the ps-ABS mode is significantly more sensitive to pairing disorder with $\delta_U=20~$nm as compared to pairing disorder with $\delta_U=40~$nm. We emphasize that this conclusion may not hold for {\em specific} disorder realizations. However, this example shows that  any stability analysis of ps-ABSs -- or, for that matter, bona fide MZMs -- should include (i) the identification of the dominant type (or types) of disorder, (ii) the characterization of the disorder profiles (e.g., the characteristic length scale $\delta_U$), and (iii) the calculation of the probability distribution of the relevant splitting susceptibilities. Such an analysis will provide bounds for the disorder strength $\delta\Lambda$ consistent with a certain acceptable energy splitting of the (near-) zero-energy mode.

\section{Summary and conclusions}\label{SecVI}

We have studied the real space and spin properties of low-energy Andreev bound states (ABSs) that emerge in semiconductor-superconductor (SM-SC) hybrid structures in the presence of potential inhomogeneities, identifying the key properties that control the collapse of these modes toward zero energy in the topologically trivial phase and determine their signatures in local charge tunneling measurements. We have established two basic scenarios for the emergence of topologically trivial low-energy ABSs corresponding to the partial separation of the component Majorana bounds states (MBSs) of these ABSs in the presence of a finite Zeeman field $\Gamma$ and i) a monotonic effective potential (i.e., a potential ``shore'') or ii) a potential hill/well. 
In order to provide an intuitive, easy-to-remember  analogy, we dubbed these two types of trivial low-energy modes ABSs generated by ($\alpha$) finite width shore potentials and ($\beta$) nearly dry potential wells/ almost submerged potential hills, respectively. In this analogy, the ``water level'' at zero magnetic field is determined by the chemical potential $\mu$, while at finite magnetic field we have a high (low) ``water level'' corresponding to $\mu$ plus (minus) half of the Zeeman splitting. Within scenario ($\alpha$), the shore width $L^*(\Gamma)$ corresponds to the  distance between the high and low water levels, while within scenario ($\beta$)  the characteristic length scale  $L^*(\Gamma)$ is given by the width of the partially ``wet bottom'' of the potential well (or the partially  ``dry top'' of the potential hill). 
We have shown that the component MBSs associated with an ($\alpha$)-type ABS belong to {\em different} spin-split sub-bands and are characterized by exponential tails pointing in the {\em same} direction, while the component MBSs associated with a ($\beta$)-type ABS belong to the {\em same} spin-split sub-band and are characterized by exponential tails pointing in {\em opposite} directions. 

We have determined that the  characteristic length scale  $L^*(\Gamma)$ associated with the effective potential controls the separation of the component MBSs, which, in turn, determines the collapse of the ABS toward zero energy. Specifically, the collapse condition within scenario  ($\alpha$) is that the separation $L^*$ of the component MBSs be larger than the characteristic width of the main peak of the MBS wave function. Note that, in general, the two partially separated MBSs (with exponential tails pointing in the same direction) have a significant overlap, but this does not lead to large energy splitting in the absence of (additional) local perturbations. 
A similar collapse condition holds for ABSs generated within scenario  ($\beta$)  by  an almost submerged potential hills (when the secondary exponential tails are suppressed and the main tails point away from each other), while  ABSs generated by nearly dry potential wells are characterized by large energy splitting oscillations due to the overlap of the exponential tails (which point toward each other), unless the MBS separation is larger than the characteristic length associated with the exponential decay of the wave function.  

We have shown that the sticking to zero energy of a trivial ABS is controlled by the real space profiles of the component MBSs (in particular by the partial separation condition), and is independent of the spin structure of the component MBSs.  In particular, we have considered a two-band model that exhibits the collapse to zero energy of an ABS consisting of component MBSs associated with the {\em same} spin sub-bands that become spatially separated in the presence of soft confinement. We also considered a `twisted' Zeeman field scenario in which two component MBSs with exponential tails pointing toward each other lead to significant energy-splitting oscillations, in spite of being associated with {\em opposite} spin sub-bands. 

Our main conclusion is that the understanding of topologically trivial ABSs emerging in systems with position-dependent effective potentials should not be based exclusively on generalizations of the soft confinement scenario, which represents a particular case of finite width shore potential. In general, the low-energy ABSs can be induced by potential inhomogeneities other than the tunnel barrier potential itself. This has implications regarding both the structure of the component MBSs (including their real-space and spin configurations) and the measurable signatures of these states in experiments involving local probes. For example, the component MBSs of an ABS induced by the soft confinement produced by a barrier potential will couple very differently to a local probe placed at the end of the wire, which results in a quantized conduction peak in a charge tunneling experiment. However, in this case the effective barrier height is not only spin-dependent, but also strongly dependent on the applied Zeeman field. A clear signature of this dependence is a strong increase of the {\em width} of the zero-bias conductance peak  with the applied Zeeman field. On the other hand, a weak dependence of the peak width on the applied Zeeman field signals that the non-homogeneity responsible for the formation of the low-energy ABS is different from the tunnel barrier itself. 

Another observable consequence of our analysis is that the emergence of well-defined energy-splitting oscillations is associated with MBSs having exponential tails pointing toward each other. These can be either MBSs localized at the ends of a short wire (with hard confinement) for $\Gamma >\Gamma_c$, or MBSs emerging locally in a nearly dry potential well in the regime $\Gamma <\Gamma_c$. By contrast, ABSs produced by soft confinement or by almost submerged potential hills are not characterized by clearly defined energy splitting oscillations. Note that the presence of quantum dots at the ends of a short wire can alter the real-space properties of the nearby MBSs and suppress the energy-splitting oscillations expected to emerge in the ``topological''  regime,  $\Gamma >\Gamma_c$. Therefore, in any experiment designed to probe the splitting oscillations it is crucial to ensure ``hard confinement.''   

We have also analyzed the stability of ps-ABSs in the presence of disorder. We find that a ps-ABS mode can be remarkably robust when  judged based on its signature in a charge tunneling experiment (i.e., in an open system), but, in essence, it is topologically unprotected. More importantly, we have proposed a quantitative scheme for analyzing the stability of Majorana modes based on calculating the probability distributions of splitting susceptibilities. Finally, we emphasize that the situations analyzed in this work illustrate the {\em basic} types of trivial low-energy ABSs that can emerge in systems with inhomogeneous effective potential, i.e. ($\alpha$)- and ($\beta$)-type ps-ABSs. In general, a trivial ps-ABS may involve a combination of the two basic scenarios. Furthermore, we have focused here on the independent band approximation, when the low-energy physics can be well captured by single-band models. More generally, in multi-band systems, inter-band couplings leading to band repulsion may also lead to the pinning of ABS modes to zero energy. Within this qualitatively different mechanism (the so-called inter-band pairing mechanism\cite{Woods2019}), the partial separation of the component MBSs does not take central stage. 

\begin{acknowledgments}
This  work  is  supported  by  NSF DMR-1414683 and ARO Grant No. W911NF-16-1-0182.
\end{acknowledgments}

\end{document}